\documentclass[12pt,epsfig]{article}  
\usepackage{amsmath}
\usepackage{amssymb}
\usepackage{graphicx}
\usepackage{subfig}
\usepackage{float}
\usepackage{commath}
\usepackage[left=2.0cm,right=2.0cm, top=2.5cm,bottom=2.5cm]{geometry}
\usepackage{cite}
\usepackage{geometry}
\usepackage{graphicx}
\usepackage{amsmath, amssymb, amsfonts}
\usepackage{hyperref}
\usepackage{cite}
\usepackage{subcaption} 
\usepackage{placeins}
\usepackage{titlesec}
\usepackage{fancyhdr}
\usepackage{caption}
\usepackage[]{geometry}
\usepackage[latin1]{inputenc}
\usepackage{commath}
\usepackage{amsmath}
\usepackage{graphicx}
\usepackage{subfig}
\usepackage{xcolor}
\usepackage{upgreek}
\usepackage{textgreek}
\usepackage{bm}
\usepackage{tabularx}
\usepackage{array}
\usepackage{xcolor} 
\usepackage{tikz-cd} 
\usepackage{multirow}
\usepackage{caption}

\begin{document}

\newcommand{\sheptitle}
{An $A_4$ model to accommodate maximal $\theta_{23}$ and maximal $\delta$ consistent 
with $\mu$--$\tau$ reflection symmetry}
\newcommand{\shepauthor}{
Rupak Chakrabarty\footnote{E-mail: rupakchakrabarty2@gmail.com},
Chandan Duarah\footnote{E-mail: chandanduarah@dibru.ac.in}
}
\newcommand{\shepaddress}
   {Department of Physics, Dibrugarh University,
               Dibrugarh - 786004, India}
\newcommand{\shepabstract}
{In this work, we construct an $A_4$-based flavor symmetry model within the framework of the type-I seesaw mechanism to realize a light neutrino mass matrix consistent with $\mu$--$\tau$ reflection symmetry. The entire framework is based on the Standard Model gauge symmetry extended by the discrete group $A_4 \times Z_2 \times Z_4$. In general, the elements of the light Majorana neutrino mass matrix are complex. The $\mu$--$\tau$ reflection symmetric texture of the mass matrix can be realized in a generalized CP symmetry limit. In this symmetry limit, the model predicts a maximal atmospheric mixing angle $\theta_{23} = \pi/4$ and a maximal Dirac CP phase $\delta = (\pi/2) / (3\pi/2)$. These features are consistent with current experimental observations, including a near-maximal value of $\theta_{23}$, a non-zero reactor angle, and a preference for $\delta$ close to $270^\circ$, as indicated by the T2K and NO$\nu$A experiments. Non-maximal values of $\theta_{23}$ and $\delta$ can be accommodated when one does not restrict to the CP symmetry limit. The model predictions for the mixing angles and the Dirac CP phase $\delta$ are then controlled by two model parameters. We perform a numerical analysis to identify the allowed values of the model parameters consistent with current global $3\nu$ oscillation data. The model successfully reproduces the desired deviations of $\theta_{23}$ and $\delta$ from their maximal values, consistent with global fit data, while simultaneously accommodating the observed values of $\theta_{12}$ and $\theta_{13}$.

\textbf{Keywords:} Lepton mixing, discrete $A_4$ symmetry,  $\mu$--$\tau$ reflection symmetry.
}
\begin{titlepage}
\begin{flushright}
\end{flushright}
\begin{center}
{\large{\bf\sheptitle}}\\[10pt]
\shepauthor\\[5pt]
{\it\shepaddress}\\[20pt]
{\bf Abstract}
\end{center}
\setcounter{page}{0}
\shepabstract
\end{titlepage}


\section{Introduction}

Neutrino oscillation experiments have firmly established the three-flavor framework of neutrino mixing, implying that neutrinos undergo flavor transitions and possess nonzero masses. This mixing phenomenon is described by the lepton mixing 
matrix (also known as PMNS matrix), a $3 \times 3$ unitary matrix that connects neutrino flavor states to their mass states. It is parameterized by three mixing angles: the solar ($\theta_{12}$), the reactor ($\theta_{13}$), and the atmospheric ($\theta_{23}$), together with one Dirac CP phase ($\delta$) and two Majorana CP phases ($\rho, \sigma$). The lepton mixing matrix is in 
general defined as:
$
U = U_\ell^\dagger U_\nu
$
where $U_\ell$ is the unitary matrix that diagonalizes the charged lepton 
mass matrix as $
U_\ell^\dagger M_\ell^\dagger M_\ell U_\ell = {M_l^\text{D}}^2= Diag(m^2_e,m^2_\mu,m^2_\tau) $ and $U_\nu$ is the unitary matrix that diagonalizes the light Majorana neutrino mass matrix as $U_\nu^\dagger M_\nu U_\nu^{*} = M_\nu^\text{D}=Diag(m_1,m_2,m_3)
$.  In the standard parametrization \cite{SN}, the lepton mixing matrix is written as
\begin{equation}
U =
\begin{pmatrix}
c_{12} c_{13} & s_{12} c_{13} & s_{13} e^{-i\delta} \\
- s_{12} c_{23} - c_{12} s_{23} s_{13} e^{i\delta}
& c_{12} c_{23} - s_{12} s_{23} s_{13} e^{i\delta}
& s_{23} c_{13} \\
s_{12} s_{23} - c_{12} c_{23} s_{13} e^{i\delta}
& - c_{12} s_{23} - s_{12} c_{23} s_{13} e^{i\delta}
& c_{23} c_{13}
\end{pmatrix}
P_{\nu}
\end{equation}
where \(c_{ij}= \cos\theta_{ij}\) and \(s_{ij}= \sin\theta_{ij}\), with \(ij = 12, 13, 23\). The diagonal matrix
$P_{\nu} = \mathrm{Diag}(e^{i\rho},\, e^{i\sigma},\, 1)$
contains the two Majorana CP phases. Since there is no experimental information on the Majorana phases, which may be probed through neutrino-less double-beta decay
$(0\nu\beta\beta)$ experiment, one may drop these phases in a particular study. In the present work we are not considering these phases in our subsequent discussions.

Neutrino oscillation data have now established the values of $\theta_{12}$ and $\theta_{13}$, with high precision\cite{Abe2025SKT2K, Abe2024SK, Allega2025SNO, Adamson2020MINOS, An2025DayaBay, Bak2018RENO, deKerret2020DoubleChooz, Abe2023T2K1, Abe2023T2K2, Himmel2020NOvA, Wolcott2024NOvA, Abbasi2025IceCube, Eguchi2003KamLAND} and it also indicate that the atmospheric mixing angle $\theta_{23}$ is nearly maximal. However, the exact position of $\theta_{23}$, whether it lies in the higher or the lower octant, is still unresolved. Recent results from the T2K and NO$\nu$A\cite{Abe2023T2K1, Abe2023T2K2, Himmel2020NOvA, Wolcott2024NOvA} experiments suggest 
that the Dirac CP phase is likely to lie near $3\pi/2$, with some 
dependence on the neutrino mass ordering. While oscillation experiments 
are insensitive to the absolute mass scale, they probe the mass-squared 
differences $\Delta m^2_{21}$ and $|\Delta m^2_{31}|$, leaving open two 
possibilities: the normal ordering ($m_1 < m_2 < m_3$) and the inverted 
ordering ($m_3 < m_1 < m_2$). Cosmological observations further constrain 
the sum of the light neutrino mass as $\sum m_i < 0.12$~eV \cite{RoyChoudhury2020}.  The values of the mixing parameters obtained from the most recent global analysis \cite{Esteban2024} are summarized in Table~\ref{tab:globalfit}.

The observed structure of the lepton
mixing matrix, in particular the approximate equality
$|U_{\mu i}| \simeq |U_{\tau i}| \quad (i=1,2,3),$ provides a strong indication of an underlying symmetry in the mixing of neutrinos.
Among the possible realizations, $\mu$--$\tau$ reflection symmetry\cite{Harrison2002} represents a
well-motivated and phenomenologically viable framework. It is basically a symmetry of lepton mixing under two combined operations,
namely the $\mu$--$\tau$ exchange operation and charge conjugation.
In terms of the lepton mixing matrix, $\mu$--$\tau$ reflection symmetry can be
defined in a mathematical form such that $U$ satisfies the condition\cite{Xing2023}
\begin{equation}
U = A_{\mu\tau}\, U^{*}\, \zeta ,
\end{equation}

where $A_{\mu\tau}$ is the $\mu$--$\tau$ exchange operator given by
\begin{equation}
A_{\mu\tau} =
\begin{pmatrix}
1 & 0 & 0 \\
0 & 0 & 1 \\
0 & 1 & 0
\end{pmatrix}.
\end{equation}
In Eq.~(2), $U^{*}$ represents the complex conjugation of $U$ and $\zeta$ is a diagonal matrix given by
$\zeta = \mathrm{Diag}(\eta_{1}, \eta_{2}, \eta_{3})$ with
$\eta_{i} = \pm 1$. The condition in Eq.~(2) leads to interesting constraints on the mixing
matrix elements:
\begin{equation}
U_{e i} = \eta_{i}\, U_{e i}^{*},
\end{equation}
and
\begin{equation}
U_{\mu i} = \eta_{i}\, U_{\tau i}^{*}.
\end{equation}

\begin{table}[t!]
\centering
\begin{tabular}{|c|c c|c c|}
\hline
 & \multicolumn{2}{c|}{Normal Ordering (NO)} & 
   \multicolumn{2}{c|}{Inverted Ordering (IO)} \\ 
\cline{2-5}
Parameter & Best-fit & $3\sigma$ range & Best-fit & $3\sigma$ range \\
\hline
\multicolumn{5}{|c|}{\textbf{Without SK}} \\
\hline
$\sin^2\theta_{12}$ & 0.307 & 0.275--0.345 & 0.308 & 0.275--0.345 \\
$\sin^2\theta_{23}$ & 0.561 & 0.430--0.596 & 0.562 & 0.437--0.597 \\
$\sin^2\theta_{13}$ & 0.02195 & 0.02023--0.02376 & 0.02224 & 0.02053--0.02397 \\
$\delta$            & 177$^\circ$ & 96--422$^\circ$ & 285$^\circ$ & 201--348$^\circ$ \\
$\Delta m^2_{21}$ ($10^{-5}\,\text{eV}^2$) & 7.49 & 6.92--8.05 & 7.49 & 6.92--8.05 \\
$\Delta m^2_{32(31)}$ ($10^{-3}\,\text{eV}^2$) 
                   & 2.534 & 2.463--2.606 & $-2.510$ & $-2.584$--$-2.438$ \\
\hline
\multicolumn{5}{|c|}{\textbf{With SK}} \\
\hline
$\sin^2\theta_{12}$ & 0.308 & 0.275--0.345 & 0.308 & 0.275--0.345 \\
$\sin^2\theta_{23}$ & 0.470 & 0.435--0.586 & 0.550 & 0.440--0.584 \\
$\sin^2\theta_{13}$ & 0.02215 & 0.02030--0.02388 & 0.02231 & 0.02060--0.02409 \\
$\delta$            & 212$^\circ$ & 214--364$^\circ$ & 274$^\circ$ & 201--335$^\circ$ \\
$\Delta m^2_{21}$ ($10^{-5}\,\text{eV}^2$) & 7.49 & 6.92--8.05 & 7.49 & 6.92--8.05 \\
$\Delta m^2_{32(31)}$ ($10^{-3}\,\text{eV}^2$) 
                   & 2.513 & 2.451--2.578 & $-2.484$ & $-2.547$--$-2.421$ \\
\hline
\end{tabular}
\caption{Best-fit values and $3\sigma$ allowed ranges of the neutrino 
oscillation parameters for NO and IO as per global analysis \cite{Esteban2024}.}
\label{tab:globalfit}
\end{table}

Eq.~(4) implies that the $e$--flavour elements $(U_{e i})$ are either purely real (when $\eta_{i}=1$) or purely imaginary (when $\eta_{i}=-1$). Further, Eq.~(5)
ensures the exact equality $|U_{\mu i}| = |U_{\tau i}|$.
Again using Eq.~(5) in the orthogonality condition
$
\sum_{i=1}^{3} U_{\mu i}\, U_{\tau i}^{*} = 0,
$
and the normalization condition
$
\sum_{i=1}^{3} \lvert U_{\mu i} \rvert^{2} = 1,
$
we can obtain the results 
\begin{equation}
\sum_{i=1}^{3} \bigl(\mathrm{Re}\,U_{\mu i}\bigr)^{2}
=
\sum_{i=1}^{3} \bigl(\mathrm{Im}\,U_{\mu i}\bigr)^{2}
=
\frac{1}{2},
\qquad
\sum_{i=1}^{3} \mathrm{Re}\,U_{\mu i}\,\mathrm{Im}\,U_{\mu i} = 0 
\label{eq:geometric_constraints}
\end{equation}
for the specific choice $\eta_{1}=\eta_{2}=\eta_{3}$.
When we substitute the elements of $U$ from the standard parametrization
in Eq.~(1) into Eq.~(6), it leads to the maximal predictions --
\begin{equation}
\theta_{23} = \frac{\pi}{4},
\qquad
\delta = \frac{\pi}{2}/\frac{3\pi}{2}.
\label{eq:maximal_predictions}
\end{equation}

The defining condition of $\mu$--$\tau$ reflection symmetry in Eq.~(2) can be
translated to the corresponding Majorana neutrino mass matrix $M_{\nu}$. Substituting
Eq.~(2) in the diagonalizing relation $M_{\nu} = U D_{\nu} U^{T},$
it can be easily shown that the Majorana mass matrix satisfies the condition
\begin{equation}
M_{\nu} = A_{\mu\tau}\, M_{\nu}^{*}\, A_{\mu\tau}.
\label{eq:mutau_condition}
\end{equation}

The mass matrix satisfying the above condition can be parametrized as
\begin{equation}
M_{\nu} =
\begin{pmatrix}
m_{ee}       & m_{e\mu}        & m_{e\mu}^{*} \\
m_{e\mu}     & m_{\mu\mu}      & m_{\mu\tau} \\
m_{e\mu}^{*} & m_{\mu\tau}     & m_{\mu\mu}^{*}
\end{pmatrix},
\end{equation}
where the elements $m_{ee}$ and $m_{\mu\tau}$ need to be real.

In recent years, $\mu$--$\tau$ reflection symmetry has been extensively
investigated as a predictive framework to understand the structure of
lepton mixing \cite{Liu2017, Nath2018, King2019, Zhao2022, Nishi2017,
Goswami2019, Zhao2017, Xing2021, Liao2020, Zhao2018, Xing2017,Duarah2021}.
This is mainly due to its noble features -- maximal $\theta_{23}$ and $\delta$ as stated in Eq.~(7), along with
a non-vanishing $\theta_{13}$. As the measurements of the T2K and NO$\nu$A experiments\cite{Abe2023T2K1, Abe2023T2K2, Himmel2020NOvA, Wolcott2024NOvA} provide a preference for $\delta$ to lie near $3\pi/2$ in inverted order scenario, it further strenghten the importance of $\mu-\tau$ reflection symmetry in lepton mixing. As far as $\mu$--$\tau$ reflection symmetry (equivalently, the
predictions of maximal $\theta_{23}$ and maximal $\delta$) is concerned, the special texture of $M_{\nu}$ in Eq.~(9) plays a crucial role in describing the lepton flavour structure.  It is also important to note that this mass-matrix texture can be realized as a consequence of radiative
correction to a degenerate neutrino mass matrix in a supersymmetric
framework~\cite{Babu2003}. The importance of this mass-matrix texture was
further discussed in Ref.~\cite{Grimus2003}. Despite these appealing
predictions, a systematic realization of this mass-matrix texture within
the framework of discrete flavor symmetries remains less explored. As for example, this mass-matrix texture has been realized in an $S_{4}$ flavor model associated with CP symmetry in Ref.~\cite{CCN2012}. Similarly, the realization of this texture in an $A_{4}$ flavor model can be found in Ref.~\cite{He2015}.

Motivated by the significant predictions of $\mu-\tau$ reflection symmetry, we attempt to construct an $A_4$ flavour model as an extension of the Standard Model (SM) gauge symmetry $SU(2)\times U(1)$, specifically to realize the $\mu$--$\tau$ reflection symmetric mass matrix texture given in Eq.~(9). The framework further incorporates an extended symmetry structure implemented through the $Z_2 \times Z_4$ symmetry, which is imposed to forbid unwanted terms in the Lagrangian. The choice of $A_4$ is motivated by its simplicity and by its ability to naturally accommodate the three generations of leptons within a single irreducible representation, leading to predictive lepton mass textures. The scalar sector of the model is extended beyond the SM by
introducing multiple $SU(2)_L$ Higgs doublets and SM gauge singlet flavon fields, following the original construction of
Ma and Rajasekaran~\cite{Ma2001} and the later formulation by
He, Keum, and Volkas~\cite{He2006}. The three left-handed lepton doublets are assigned to a triplet representation of the $A_4$ symmetry, while the right-handed charged leptons transform as distinct singlets. In addition, three right-handed neutrinos are introduced, allowing the implementation of the Type-I seesaw mechanism and the generation of light neutrino masses. The interplay between the extended scalar sector and the imposed symmetries results in a constrained structure for both the Dirac and Majorana mass matrices, which leads to the desired neutrino mass structure given in Eq.~(9).

In general, the elements of the effective light neutrino mass matrix
obtained via the seesaw mechanism are complex. In the present model, the complex light neutrino mass matrix is expressed in the flavour basis where the charged lepton mass matrix is diagonal. This transformation allows us to obtain
the mass matrix texture associated with $\mu$--$\tau$ reflection symmetry. However, to obtain the exact form of the neutrino mass matrix given in Eq.~(9), we impose a generalized CP symmetry at the
level of the model Lagrangian. This symmetry requires the
Yukawa couplings and the vacuum expectation values to be real. As a consequence, the mass-matrix parameters become real. In addition, an equality condition between two coupling constants in the charged-lepton sector is imposed. Together, these conditions ensure the realization of the $\mu$--$\tau$ reflection--symmetric neutrino mass matrix of the form given in Eq.~(9). In this mass matrix, the complex structure arises solely from the Clebsch--Gordan coefficients of the $A_4$ flavor group. Once the $\mu$--$\tau$ reflection--symmetric neutrino mass matrix is established, its phenomenological consequences can be systematically analyzed. The resulting PMNS matrix exhibits several characteristic features, including maximal atmospheric mixing and a maximally CP-violating Dirac phase. These predictions arise as direct consequences of the underlying $\mu$--$\tau$ reflection symmetry and are independent of detailed parameter choices. In the subsequent sections, we generalize this framework to the complex case, where nontrivial phases enter through the Clebsch--Gordan coefficients of the $A_4$ symmetry, and study how the mixing angles and CP phase are modified. A detailed numerical analysis is then performed to confront the model predictions with current neutrino oscillation data for both normal and inverted mass orderings.

The rest of the paper is organized as follows: in Section~2, we describe the basic structure of the model, including the field content and
their transformation properties under the imposed symmetries. We construct the Yukawa Lagrangian and derive both the Dirac and Majorana neutrino mass matrices. The effective light neutrino mass matrix is then obtained via the Type-I seesaw mechanism, and its connection to the realization of $\mu$--$\tau$ reflection symmetry is discussed. In Section~3, the full scalar potential is constructed, and a detailed analysis of its minimization is carried out to demonstrate how the required
vacuum expectation value alignments arise consistently within the symmetry framework. In Section~4, we generalize the analysis to the complex case, allowing deviations from exact $\mu$--$\tau$ reflection symmetry. We derive the modified expressions for the lepton mixing angles and the Dirac CP phase, and perform a comprehensive numerical analysis to determine the allowed regions of the model parameters consistent with current neutrino oscillation data for both normal and inverted mass orderings. We conclude with a summary of the main results in the final section.

\section{Basic structure of the model}

Our framework is constructed on the basis of the SM gauge group $SU(2) \times U(1)$. In this framework, we incorporate the discrete flavor symmetry $A_4$, which is widely used in flavor model building due to its simple group structure and the ability to generate realistic lepton mixing patterns\cite{Brahmachari2008, King2012, Altarelli2010, Babu2003, Grimus2003, CCN2012, He2015, Ma2001, He2006, Altarelli2005, Altarelli2006, Ma2004,Ahn2012,Karmakar2015}. In addition, the auxiliary $Z_2 \times Z_4$ symmetries are imposed to forbid unwanted couplings. As a result, the full symmetry of the model is 
$SU(2) \times U(1) \times A_4 \times Z_2 \times Z_4$. For completeness, the properties and product representations of the $A_4$ group are briefly summarized in Appendix~A.

The $A_4$ flavour structure of our model is primarily based on the original framework introduced by Ma and Rajasekaran~\cite{Ma2001}, as well as a later model by He, Keum, and Volkas~\cite{He2006}.
Accordingly, the left-handed lepton doublets
$l_L = (l_L^1,l_L^2,l_L^3)$ are assigned to the $A_4$ triplet,
while the right-handed charged leptons $l_R^1$, $l_R^2$, and
$l_R^3$ transform as the $A_4$ singlets $\mathbf{1}$, $\mathbf{1}''$, and $\mathbf{1}'$, respectively. Three right-handed neutrinos $N_R = (N_R^1,N_R^2,N_R^3)$ are also added as $A_4$ triplets, which enables the Type-I seesaw mechanism to generate the light Majorana neutrino masses. In the original model introduced by Ma and Rajasekaran\cite{Ma2001},  the scalar sector consists of four $SU(2)$ doublets -- three of which form an $A_4$ triplet and the remaining one transforms as $A_4$ singlet. In the present model
we consider two additional $SU(2)$ doublet scalar fields transforming as $A_4$ singlets. In total there are six $SU(2)$
doublet scalar fields in our model out of which three, $\Phi$=
($\Phi_1$,$\Phi_2$,$\Phi_3$) transform as $A_4$ triplet and others 
$\eta_1$,$\eta_2$ and $\eta_3$ transform as $\mathbf{1}$, $\mathbf{1}''$, and $\mathbf{1}'$ respectively under $A_4$. 
Further, similar to the model by He, Keum, and  
Volkas~\cite{He2006}, the present framework also includes  
three flavon fields $\chi = (\chi_1, \chi_2, \chi_3)$,  
which are SM singlets and transform as triplet under $A_4$. The field content and
symmetry assignments of the model are summarized in Table~2.

\begin{table}[t]
    \centering
    \small 
    \begin{tabular}{|c|c|c|c|c|c|c|c|c|c|c|}
        \hline
        & $l_L$ & $l_R^1$ & $l_R^2$ & $l_R^3$ & $N_R^i$ & $\Phi$ & $\chi$ & $\eta_1$ & $\eta_2$ & $\eta_3$ \\  
        \hline
        SU(2)$_L$ & 2 & 1 & 1 & 1 & 1 & 2 & 1 & 2 & 2 & 2 \\  
        \hline
        A$_4$ & 3 & 1 & $1''$ & $1'$ & 3 & 3 & 3 & 1 & $1'$ & $1''$ \\  
        \hline
        Z$_2$ & + & + & + & + & -- & + & + & -- & -- & -- \\  
        \hline
        Z$_4$ & $\omega$ & 1 & 1 & 1 & $\omega^2$ & $\omega^3$ & 1 & $\omega$ & $\omega$ & $\omega$ \\  
        \hline
    \end{tabular}
    \caption{The field content and transformation properties under the imposed symmetries.}
\end{table}
The Yukawa lagrangian of the lepton sector invariant under the full
gauge group $SU(2) \times U(1) \times A_{4} \times Z_{2} \times Z_{4}$ is given by
 \begin{align}
\mathcal{L}_{Y} \;=\;&\ 
y_{e}\,(\bar{l}_{L}\,\Phi)_{\mathbf{1}}\,l^{1}_{R}
+ y_{\mu}\,(\bar{l}_{L}\,\Phi)_{\mathbf{1}'}\,l^{2}_{R}
+ y_{\tau}\,(\bar{l}_{L}\,\Phi)_{\mathbf{1}''}\,l^{3}_{R} \notag\\
&+ \lambda^{1}_{N}\,(\bar{l}_{L}\,N_{R})_{\mathbf{1}}\,\eta_{1}
+ \lambda^{2}_{N}\,(\bar{l}_{L}\,N_{R})_{\mathbf{1}''}\,\eta_{2}
+ \lambda^{3}_{N}\,(\bar{l}_{L}\,N_{R})_{\mathbf{1}'}\,\eta_{3} \notag\\
&+ m\,(\bar{N}_{R}\,N^{C}_{R})_{\mathbf{1}}
+ \lambda_{\chi}\,(\bar{N}_{R}\,N^{C}_{R})_{\mathbf{3}}\,\chi
+ \text{h.c.}
\end{align}

The first row on the RHS of Eq.~(10) corresponds to the charged-lepton sector and represents the Yukawa interactions of the lepton doublets with the Higgs field $\Phi$. When $\Phi$ acquires a vacuum expectation value (vev) given by
\begin{equation}
\langle \Phi \rangle = \langle \Phi_1, \Phi_2, \Phi_3 \rangle = (v_\Phi,\, v_\Phi,\, v_\Phi).
\end{equation}
After symmetry breaking, these Yukawa interactions yield the charged-lepton mass matrix given by
\begin{equation}
M_{\ell} =
\begin{pmatrix}
y_e v_{\Phi} & y_\mu v_{\Phi} & y_\tau v_{\Phi} \\[1mm]
y_e v_{\Phi} & \omega\, y_\mu v_{\Phi} & \omega^{2} y_\tau v_{\Phi} \\[1mm]
y_e v_{\Phi} & \omega^{2} y_\mu v_{\Phi} & \omega\, y_\tau v_{\Phi}
\end{pmatrix}, \qquad \omega = e^{2\pi i/3}.
\end{equation}
  
This mass matrix is diagonalized through a unitary transformation $U_\ell$ acting on the left-handed fields and a transformation $U_R$ acting on right handed fields:
\begin{equation}
U_{\ell}^{\dagger}\, M_{\ell}\, U_{R} = M_{\ell}^{\rm diag},
\end{equation}
where $U_{\ell}$ is given by
\begin{equation}
U_{\ell} = \frac{1}{\sqrt{3}}
\begin{pmatrix}
1 & 1 & 1 \\[1mm]
1 & \omega & \omega^{2} \\[1mm]
1 & \omega^{2} & \omega
\end{pmatrix},
\end{equation}
and $U_{R}$ is simply the identity matrix.  
The resulting diagonal mass matrix is   
 
\begin{equation}
M_{\ell}^{\rm diag} =
\begin{pmatrix} 
\sqrt{3}\, y_e v_{\Phi} & 0 & 0 \\[1mm]
0 & \sqrt{3}\, y_\mu v_{\Phi} & 0 \\[1mm]
0 & 0 & \sqrt{3}\, y_\tau v_{\Phi} 
\end{pmatrix}
\end{equation}  
which provides the physical charged-lepton masses
$
m_e = \sqrt{3}\, y_e v_{\Phi},\
m_\mu = \sqrt{3}\, y_\mu v_{\Phi},\
m_\tau = \sqrt{3}\, y_\tau v_{\Phi}.
$

The neutrino sector consists of two parts: one generating the Dirac neutrino mass and the other corresponding to the heavy Majorana neutrino mass. The Yukawa terms in the second row on the RHS of Eq.~(10) are responsible for generating the Dirac neutrino masses. With the specific vacuum expectation value alignment
\begin{equation}
\langle \eta_1 \rangle = \langle \eta_2 \rangle = \langle \eta_3 \rangle = v_{\eta},
\end{equation}
we get the Dirac neutrino mass matrix as
\begin{equation}
M_D =
\begin{pmatrix} 
\lambda^1_N v_\eta+\lambda^2_N v_\eta+\lambda^3_N v_\eta & 0 & 0 \\[1mm]
0 & \lambda^1_N v_\eta+\omega^2\lambda^2_N v_\eta+\omega\lambda^3_N v_\eta & 0 \\[1mm]
0 & 0 & \lambda^1_N v_\eta+\omega\lambda^2_N v_\eta+\omega^2\lambda^3_N v_\eta
\end{pmatrix}. 
\end{equation}
It is important to note that the coupling term $(\bar{l}_L N_R) \Phi$, although allowed by the $A_4$ symmetry, could have contributed to the Dirac neutrino mass $M_D$. However, this term is simultaneously forbidden by the $Z_2$ and $Z_4$ symmetries. Its absence is therefore crucial for maintaining the desired structure of $M_D$.

The heavy Majorana mass for the right-handed neutrinos receives contributions from two sources. 
The first one is a bare mass term $m\,(\bar{N}_{R}\,N^{C}_{R})_{\mathbf{1}}$
in Eq.~(10), providing a uniform contribution to all three right-handed neutrinos in diagonal form. 
The second arises from the interaction of the right-handed neutrinos, $N_R$ with the flavon triplet $\chi$. 
After the flavons acquire the vev pattern
\begin{equation}
\langle \chi \rangle = \langle \chi_1, \chi_2, \chi_3 \rangle = (0,\, v_\chi,\, 0),
\end{equation}
the last interaction term in Eq.~(10) generates off-diagonal entries in the Majorana mass matrix. Thus, the resulting heavy Majorana mass matrix is given by
\begin{equation}
M_R =
\begin{pmatrix}
m & 0 & \lambda_\chi v_\chi \\[1mm]
0 & m & 0 \\[1mm]
\lambda_\chi v_\chi & 0 & m
\end{pmatrix}.
\end{equation}

Finally the light effective Majorana neutrino mass matrix is generated via the Type-I seesaw formula:  
\begin{equation}
M_\nu = - M_D \, M_R^{-1} \, M_D^T =
\begin{pmatrix}
M_\nu^{11} & 0 & M_\nu^{13} \\[1mm]
0 & M_\nu^{22} & 0 \\[1mm]
M_\nu^{13} & 0 & M_\nu^{33}
\end{pmatrix}
\end{equation}
  
with the elements given by
\begin{align}
M_\nu^{11} &=
- \frac{m\, v_\eta^{2}}{m^{2} - (\lambda_\chi v_\chi)^{2}}
\left(
\lambda_N^{1} + \lambda_N^{2} + \lambda_N^{3}
\right)^{2}, \notag \\[1mm]
M_\nu^{13} &=
- \frac{(\lambda_\chi v_\chi)^{2}\, v_\eta^{2}}{m^{2} - (\lambda_\chi v_\chi)^{2}}
\left(
\lambda_N^{1} + \lambda_N^{2} + \lambda_N^{3}
\right)
\left(
\lambda_N^{1}
+ \lambda_N^{2}\,\omega
+ \lambda_N^{3}\,\omega^{2}
\right), \notag \\[1mm]
M_\nu^{22} &=
- \frac{v_\eta^{2}}{m}
\left(
\lambda_N^{1} + \lambda_N^{2} + \lambda_N^{3}
\right)
\left(
\lambda_N^{1}
+ \lambda_N^{2}\,\omega^{2}
+ \lambda_N^{3}\,\omega
\right), \notag \\[1mm]
M_\nu^{33} &=
- \frac{m\, v_\eta^{2}}{m^{2} - (\lambda_\chi v_\chi)^{2}}
\left(
\lambda_N^{1}
+ \lambda_N^{2}\,\omega
+ \lambda_N^{3}\,\omega^{2}
\right)^{2}.
\end{align}

In the flavor basis where the charged-lepton mass matrix is diagonal
(Eq.~(14)), the light neutrino mass matrix in Eq.~(20) becomes
\begin{equation}
\begin{aligned}
M'_\nu &= U_\ell^\dagger \, M_\nu \, U_\ell^\ast \\
&=
\resizebox{0.98\textwidth}{!}{$
\begin{pmatrix}
M_\nu^{11} + 2 M_\nu^{13} + M_\nu^{22} + M_\nu^{33}
&
M_\nu^{11} - \omega^2 M_\nu^{13} + \omega^2 M_\nu^{22} + \omega M_\nu^{33}
&
M_\nu^{11} - \omega M_\nu^{13} + \omega M_\nu^{22} + \omega^2 M_\nu^{33}
\\[1mm]
M_\nu^{11} - \omega^2 M_\nu^{13} + \omega^2 M_\nu^{22} + \omega M_\nu^{33}
&
M_\nu^{11} + 2 \omega M_\nu^{13} + \omega M_\nu^{22} + \omega^2 M_\nu^{33}
&
M_\nu^{11} - M_\nu^{13} + M_\nu^{22} + M_\nu^{33}
\\[1mm]
M_\nu^{11} - \omega M_\nu^{13} + \omega M_\nu^{22} + \omega^2 M_\nu^{33}
&
M_\nu^{11} - M_\nu^{13} + M_\nu^{22} + M_\nu^{33}
&
M_\nu^{11} + 2 \omega^2 M_\nu^{13} + \omega^2 M_\nu^{22} + \omega M_\nu^{33}
\end{pmatrix}
$}
\end{aligned}
\end{equation}

The above mass matrix fulfills the central goal of the present work. In general, the parameters $M_\nu^{11}$, $M_\nu^{13}$,
$M_\nu^{22}$, and $M_\nu^{33}$ are complex, and the mass matrix is
complex symmetric, in consistent with the Majorana nature of neutrinos. These parameters can receive complex phases from different sources, including the Yukawa couplings, the vacuum expectation values (vevs) of scalar fields, or group-theoretical factors such as the roots of unity $\omega_i$ appearing in the mass terms. The striking property of this mass matrix is that it carries the texture of the $\mu$--$\tau$ reflection--symmetric mass matrix presented in Eq.~(9), if we impose all the elements of the mass matrix $M_\nu$ to be real. When all these elements become real it immediately follows from Eq.~(22) that $(M'_\nu)_{11}$ and $(M'_\nu)_{23}$ are real, with $(M'_\nu)_{12}^{*} = (M'_\nu)_{13}$ and $(M'_\nu)_{22}^{*} = (M'_\nu)_{33}$. In this work, we restrict ourselves to the real nature of the elements of $M_\nu$ such that the mass matrix $M'_\nu$ in Eq.~(22) assumes the $\mu$-$\tau$ reflection symmetric texture given in Eq.~(9).

The real nature of the elements $M_\nu^{11}$, $M_\nu^{13}$, $M_\nu^{22}$, and $M_\nu^{33}$ can be realized by invoking a generalized CP symmetry~\cite{Feruglio2013,Holthausen2013,King2014,Nishi2016,Ahn2013,Ding2013} on the Yukawa Lagrangian in Eq.~(10), along with the special condition $\lambda_N^{2} = \lambda_N^{3}$. The generalized CP symmetry generally refers to a symmetry under the conventional CP transformation of given fields accompanied with a nontrivial
permutation of the flavor indices, corresponding to the interchange of the second and third generations ($2 \leftrightarrow 3$).The latter acts on the fields that transform as $A_4$ triplets, namely the left-handed lepton doublets $l_L$, the right-handed neutrinos $N_R^i$, and the scalar fields $\Phi$ and $\chi$. Thus, the transformation of the $A_4$ triplets under the generalized CP transformation may be expressed as
\begin{align*}  
(l^1_L,\, l^2_L,\, l^3_L) &\rightarrow \left((l^1_L)^{CP},\, (l^3_L)^{CP},\, (l^2_L)^{CP}\right), \\  
(N^1_R,\, N^2_R,\, N^3_R) &\rightarrow \left((N^1_R)^{CP},\, (N^3_R)^{CP},\, (N^2_R)^{CP}\right), \\  
(\Phi_1,\, \Phi_2,\, \Phi_3) &\rightarrow \left(\Phi_1^\dagger,\, \Phi_3^\dagger,\, \Phi_2^\dagger\right), \\  
(\chi_1,\, \chi_2,\, \chi_3) &\rightarrow \left(\chi_1^\dagger,\, \chi_3^\dagger,\, \chi_2^\dagger\right).  
\end{align*}  
The singlet fields simply transform under the usual CP symmetry as  
\[
l_R^i \rightarrow (l_R^i)^{CP}, \quad \eta_i \rightarrow \eta_i^\dagger,
\]  
where $i = 1,2,3$. Under such generalized CP transformations, the Yukawa Lagrangian in Eq.~(10) remains invariant, which implies that all Yukawa couplings and scalar vacuum expectation values are real. This, in turn, ensures that the elements of $M_\nu$ become real, as is evident from Eq.~(21) when the special condition $\lambda_N^{2} = \lambda_N^{3}$ is imposed. It is worth noting that, with all Yukawa couplings and vacuum expectation values being real, the only remaining source of CP violation arises from the complex Clebsch-Gordan coefficients of the $A_4$ flavor group.

It is important to note that a similar approach in obtaining a light
Majorana mass matrix having a $\mu$--$\tau$ reflection symmetric texture is also made by X-G He in Ref.~\cite{He2015}. A primary difference between~\cite{He2015} and our approach exists regarding the introduction of the three scalar fields $\eta_1$, $\eta_2$, and $\eta_3$, which transform as $\mathbf{1}$, $\mathbf{1}''$, and $\mathbf{1}'$, respectively, under $A_4$. In~\cite{He2015}, two such scalar fields are considered as SM singlets, while in this work we have considered all the three scalar fields as SM doublets in a uniform manner.

With the realization of a mass matrix having the texture of
$\mu$--$\tau$ reflection symmetry, let us now turn to the discussion of the corresponding lepton mixing matrix. With all the parameters being real, the neutrino mass matrix $M_\nu$ in Eq.~(20) is real and symmetric, allowing diagonalization by a real orthogonal matrix $U_\nu^r$ as:

\begin{equation}
M_\nu^{\text{diag}} = (U_\nu^{r})^T M_\nu U_\nu^{r}
\end{equation}
where
\begin{equation}
U_\nu^r =
\begin{pmatrix}
\cos\theta & 0 & -\sin\theta \\
0 & 1 & 0 \\
\sin\theta & 0 & \cos\theta
\end{pmatrix}.
\end{equation}

Then the corresponding lepton mixing matrix is given by
\begin{equation}
U^{\mu\tau} = U_\ell^\dagger U_\nu^r = \frac{1}{\sqrt{3}}
\begin{pmatrix}
\cos\theta + \sin\theta & 1 & \cos\theta - \sin\theta \\
\cos\theta + \omega \sin\theta & \omega^2 & \omega\cos\theta - \sin\theta \\
\cos\theta + \omega^2 \sin\theta & \omega & \omega^2\cos\theta - \sin\theta
\end{pmatrix},
\end{equation}
where $U_\ell$ is given in Eq.~(14).From the above lepton mixing matrix, the mixing angles defined in the standard parametrization (Eq.~(1)) can be obtained as

\begin{equation}
\sin^{2}\theta_{12}
=
\frac{|U_{e2}|^{2}}{1 - |U_{e3}|^{2}}
=
\frac{1}{2 + \sin 2\theta} ,
\label{eq:theta12}
\end{equation}

\begin{equation}
\sin^{2}\theta_{23}
=
\frac{|U_{\mu 3}|^{2}}{1 - |U_{e3}|^{2}}
=
\frac{1}{2} ,
\label{eq:theta23}
\end{equation}

\begin{equation}
\sin^{2}\theta_{13}
=
|U_{e3}|^{2}
=
\frac{1 - \sin 2\theta}{3} .
\label{eq:theta13}
\end{equation}

Thus, we have arrived at the predictions of maximal atmospheric mixing (Eq.~(27)), and a nonzero $\theta_{13}$ (Eq.~(28)), as promised by $\mu$--$\tau$ reflection symmetry. As reflection symmetry also predicts a maximal value of the Dirac CP phase $\delta$, it can be explicitly verified from the lepton mixing matrix in Eq.~(25) using the Jarlskog invariant. The Jarlskog invariant, defined as
$
J = \mathrm{Im}\!\left(
U_{e1} U_{e2}^* U_{\mu1}^* U_{\mu2}
\right),
$
corresponding to the lepton mixing matrix in the standard
parametrization in Eq.~(1), is given by:

\begin{equation}
J = \, \sin\theta_{12} \, \cos\theta_{12} \,
\sin\theta_{23} \, \cos\theta_{23} \, \sin\theta_{13} \,\cos^2\theta_{13}  \sin\delta.
\end{equation}

On the other hand, the Jarlskog invariant corresponding to the lepton mixing matrix in Eq.~(25) can be calculated as
\begin{equation}
J = -\frac{1}{6\sqrt{3}} \cos(2\theta).
\end{equation}

By equating Eqs.~(29) and (30) and using Eqs.~(26)-(28), one finds that
\begin{equation}
\delta = 
\begin{cases}
\dfrac{3\pi}{2}, & \text{if } \cos(2\theta) > 0, \\[4pt]
\dfrac{\pi}{2}, & \text{if } \cos(2\theta) < 0.
\end{cases}
\end{equation}

It is evident that the Jarlskog invariant $J$ is nonzero, 
demonstrating intrinsic CP violation arising from the complex 
structure of the Clebsch--Gordan coefficients in the $A_4$ symmetry. 
Together with maximal $\theta_{23}$, this highlights the predictive 
power of the $\mu$--$\tau$ reflection symmetry in our model.

\section{Scalar Potential}

The total scalar potential for the model can be written as
\begin{equation}
V = V(\Phi) + V(\chi) + V(\eta_i) + V(\Phi, \chi) + V(\eta_i, \eta_j) + V(\Phi, \eta_i) + V(\Phi, \eta_i, \eta_j)
+ V(\eta_i, \chi) + V(\eta_i, \eta_j, \chi) + V(\Phi, \chi, \eta_i),
\end{equation}
where the terms on the right-hand side represent the corresponding contributions arising from the relevant self and mutual interactions of the scalar fields. The self-interaction terms corresponding to the Higgs triplet $\Phi$ and the scalar triplet $\chi$ are given by
\begin{equation}
\begin{aligned}[t]
V(\Phi) = \lambda_1^{\Phi} (\Phi^\dagger \Phi)_1 (\Phi^\dagger \Phi)_1 
+ \lambda_2^{\Phi} (\Phi^\dagger \Phi)_{1'} (\Phi^\dagger \Phi)_{1''} + \lambda_3^{\Phi} (\Phi^\dagger \Phi)_{3s} (\Phi^\dagger \Phi)_{3s} 
+ \lambda_4^{\Phi} (\Phi^\dagger \Phi)_{3a} (\Phi^\dagger \Phi)_{3a}\\
+ \left[\lambda_5^{\Phi} (\Phi^\dagger \Phi)_{3s} (\Phi^\dagger \Phi)_{3a} + \text{H.C.}\right].
\end{aligned}
\end{equation}

\begin{equation}
\begin{aligned}[t]
V(\chi) = \mu^2_\chi (\chi \chi)_1 
+ \lambda^\chi_1 (\chi \chi)_1 (\chi \chi)_1
+ \lambda^\chi_2 (\chi \chi)_{1'} (\chi \chi)_{1''} + \lambda^{\chi}_3 (\chi \chi)_{3_s} (\chi \chi)_{3_s}
+ \lambda^{\chi}_4 (\chi \chi)_{3_a} (\chi \chi)_{3_a}\\ + \lambda^{\chi}_5 (\chi \chi)_{3_s} (\chi \chi)_{3_a}
+  \xi^{\chi}_1 \chi (\chi \chi)_{3_s}
+ \xi^{\chi}_2 \chi (\chi \chi)_{3_a}.
\end{aligned}
\end{equation}
The potential terms corresponding to the mutual interaction between triplet fields $\Phi$ and $\chi$ are given by 
\begin{equation}
\begin{aligned}
V(\Phi, \chi) = \lambda_1^{\Phi\chi} (\Phi^\dagger \Phi)_1 (\chi^\dagger \chi)_1 
+ \lambda_2^{\Phi\chi} (\Phi^\dagger \Phi)_{1'} (\chi^\dagger \chi)_{1''} + \lambda_3^{\Phi\chi} (\Phi^\dagger \Phi)_{1''} (\chi^\dagger \chi)_{1'} 
+ \lambda_4^{\Phi\chi} (\Phi^\dagger \Phi)_{3s} (\chi^\dagger \chi)_{3a} \\
+ \lambda_5^{\Phi\chi} (\Phi^\dagger \Phi)_{3a} (\chi^\dagger \chi)_{3s}
+ \lambda_6^{\Phi\chi} (\Phi^\dagger \Phi)_{3s} \chi 
+ \lambda_7^{\Phi\chi} (\Phi^\dagger \Phi)_{3a} \chi.
\end{aligned}
\end{equation}
The self and mutual interaction terms involving the scalar singlets $\eta_i$ are
\begin{equation}
V(\eta_1) = \lambda^{\eta_1} (\eta_1^\dagger \eta_1)^2, 
\end{equation}
\begin{equation}
V(\eta_1, \eta_2) = \lambda_3^{\eta_1 \eta_2} (\eta_2^\dagger \eta_2)(\eta_1^\dagger \eta_2) + \text{H.C.},
\end{equation}
\begin{equation}
V(\eta_1, \eta_3) = \lambda_3^{\eta_1 \eta_3} (\eta_3^\dagger \eta_3)(\eta_1^\dagger \eta_3) + \text{H.C.},
\end{equation}
\begin{equation}
V(\eta_2, \eta_3) = [\lambda^{\eta_2 \eta_3} (\eta_2^\dagger \eta_3)^2 + \text{H.C.}]
+ \lambda_1^{\eta_2 \eta_3} (\eta_2^\dagger \eta_2)(\eta_3^\dagger \eta_3) + \lambda_2^{\eta_2 \eta_3} (\eta_2^\dagger \eta_3)(\eta_3^\dagger \eta_2).
\end{equation}
Finally, the interaction terms between $\Phi$ and $\eta_i$ fields are
\begin{equation}
V(\Phi, \eta_1) = \lambda^{\Phi \eta_1} (\Phi^\dagger \Phi)_1 (\eta_1^\dagger \eta_1),
\end{equation}
\begin{equation}
V(\Phi, \eta_2, \eta_3) = \lambda^{\Phi \eta_{23}} (\Phi^\dagger \Phi)_1 (\eta_2^\dagger \eta_3) + \text{H.C.}.
\end{equation}

In the above expressions $\mu_\chi$, $\xi^{\chi}_1$, and $\xi^{\chi}_2$ have mass  
dimension one, whereas all other coupling constants remain dimensionless.  
It is important to note that the terms in $V(\Phi, \chi)$ given in Eq.~(35), arising from the interactions between $\Phi$ and $\chi$, 
creates a serious problem in obtaining vacuum expectation value (vev) solutions \cite{He2006}. This is because the minimization of the scalar potential yields more independent equations than the number of vevs, namely $v_{\Phi}$, $v_{\chi}$, and $v_{\eta}$, making it difficult to find a solution. One effective way to avoid this is to eliminate  the interaction terms present in $V(\Phi, \chi)$. This  can be naturally realized by imposing a $Z_4$ symmetry,  
under which all such interaction terms are automatically removed.  
Along similar lines, no renormalizable term involving $\Phi$, $\eta$, and  
$\chi$ simultaneously is permitted by the $Z_2$ symmetry.  
As a result, potential terms like $V(\Phi, \chi, \eta_i)$ 
do not contribute to the total potential of the model. It is important to note that the self-interaction terms $V(\eta_2)$ and $V(\eta_3)$ are forbidden by the $A_4 \times Z_4$ symmetry. In contrast, for $V(\eta_1)$, the only allowed term, consistent with the same symmetry, is the quartic interaction $\lambda^{\eta_1} (\eta_1^\dagger \eta_1)^2$, as given in Eq.~(36).  
Regarding the mutual interaction terms arising from interactions among $\eta_i$ and $\eta_j$ (with $i \neq j$), only the terms given in Eqs.~(37)-(39) are allowed by the $A_4$ symmetry, while all other terms are forbidden.  
Among the interaction terms $V(\Phi, \eta_i)$ for $i = 1, 2, 3$, only the term involving $\eta_1$, given in Eq.~(40), is allowed by the $A_4$ symmetry. The interactions with $\eta_2$ and $\eta_3$ are forbidden, as they do not form singlet combinations under $A_4$ and are therefore excluded from the potential. In the same spirit, among the mutual interaction terms involving $\Phi$ and two distinct $\eta_i$ fields, only $V(\Phi, \eta_2, \eta_3)$, given in Eq.~(41), is allowed, while all others are forbidden by the $A_4$ symmetry. Furthermore, all interactions of the form $V(\eta_i, \chi)$ and $V(\eta_i, \eta_j, \chi)$ (with $i \neq j$) are forbidden by both $A_4$ and $Z_4$ symmetries.

Notably, the term $V(\Phi)$ in the present model 
does not include a quadratic mass term of the form $\mu^2_\Phi (\Phi^\dagger \Phi)$, 
as such a term is forbidden by the imposed $\mathbb{Z}_4$ symmetry. This quadratic 
term typically plays an important role in inducing the vev of 
$\Phi$ in standard scenarios. In our framework, however, the vev of $\Phi$ is generated 
dynamically through its interactions with the auxiliary scalar fields $\eta_i$, 
specifically via the mixed terms in $V(\Phi, \eta_i)$ and $V(\Phi, \eta_i, \eta_j)$. 
These interactions effectively drive spontaneous symmetry breaking, eliminating the 
need for an explicit mass parameter in $V(\Phi)$.Now, the minimization of the scalar potential with respect to $\Phi_1^*$ gives
\begin{align}
\left. \frac{\partial V}{\partial \Phi_1^*} \right|_{\langle \Phi_i \rangle = v_{\Phi},\, \langle \eta_i \rangle = v_{\eta}} 
&= 2 \lambda_1^{\Phi} v_{\phi} \left( |v_{\phi}|^2 + |v_{\phi}|^2 + |v_{\phi}|^2 \right) \nonumber \\
&\quad + 2 \lambda_3^{\Phi} \left[ 
\left( v_{\Phi}^* v_{\Phi} + v_{\Phi}^* v_{\Phi} \right) v_{\Phi} 
+ \left( v_{\Phi}^* v_{\Phi} + v_{\Phi}^* v_{\Phi} \right) v_{\Phi} 
\right] \nonumber \\
&\quad + \lambda_1^{\Phi \eta_1} v_{\Phi} |v_{\eta}|^2
+ \lambda^{\Phi \eta_{23}} v_{\phi} |v_{\eta}|^2 
+ \lambda^{\Phi \eta_{23}*} v_{\phi} |v_{\eta}|^2 = 0.
\end{align}
This equation gives the expression of the vev of $\Phi$ as
\begin{equation}
v_{\Phi} = \sqrt{ 
\frac{ 
- \left( \lambda_1^{\Phi \eta_1} + \lambda^{\Phi \eta_{23}} + \lambda^{\Phi \eta_{23}*} \right) |v_{\eta}|^2 
}{
6 \lambda_1^{\Phi} + 4 \lambda_3^{\Phi}
} }.
\end{equation}
Similarly, minimizing the scalar potential with respect to $\chi_2$ yields
\begin{equation}
\left. \frac{\partial V}{\partial \chi_2} \right|_{\langle \chi_i \rangle = (0, v_{\chi_2}, 0)}= v_{\chi_2} \left( 2 \mu_\chi^2 + 4 \lambda_1^{\chi} v_{\chi_2}^2 
+ 4 \lambda_2^{\chi} v_{\chi_2}^2 \right)= 0,
\end{equation}
which leads to the vev of $\chi$ as
\begin{equation}
v_{\chi_2} = \sqrt{ \frac{ -\mu_\chi^2 }{ 2(\lambda_1^{\chi} + \lambda_2^{\chi}) } }.
\end{equation}
Now, to obtain the vev of $\eta$, we minimise the scalar potential with respect to $\eta_1^*$, which yields
\begin{equation}
\left. \frac{\partial V}{\partial \eta_1^*} \right|_{\langle \eta_1, \eta_2, \eta_3 \rangle = (v_{\eta}, v_{\eta}, v_{\eta})} = 2 \lambda^{\eta_1} |v_\eta|^2 v_\eta 
+ 3 \lambda_1^{\Phi \eta_1} |v_\phi|^2 v_\eta 
+ \lambda_3^{\eta_1 \eta_2} |v_\eta|^2 v_\eta 
+ \lambda_3^{\eta_1 \eta_3} |v_\eta|^2 v_\eta =0.
\end{equation}
From this equation, the vev of $\eta$ is obtained as
\begin{equation}
v_{\eta} = \sqrt{ 
\frac{ 
-3 \lambda_1^{\Phi \eta_1} |v_{\phi}|^2 
}{
2 \lambda^{\eta_1} + \lambda_3^{\eta_1 \eta_2} + \lambda_3^{\eta_1 \eta_3}
} }.
\end{equation}

From the above expression of $v_{\eta}$, it is clear that no term of the form 
\( \mu_{\eta_i}^2 \) contributes to the generation of \( \langle \eta_i \rangle \), 
as a quadratic term like \( \mu_{\eta_i}^2 (\eta_i^\dagger \eta_i) \) is forbidden 
in the scalar potential due to the imposed \( A_4 \times Z_4 \) symmetry. Consequently, 
the vacuum expectation value of \( \eta_i \) arises solely from cross-interaction 
terms involving \( \Phi \) and \( \eta_i \), particularly those proportional to 
\( \lambda_1^{\Phi \eta_1} \). These interactions effectively induce spontaneous symmetry 
breaking, underscoring the dynamical origin of the vev in the absence of a bare mass 
parameter for \( \eta_i \).

\section{Scenario of non-maximal $\theta_{23}$ and non-maximal $\delta$}

In order to allow non-maximal $\theta_{23}$ and non-maximal $\delta$, we return to the general situation where the mass matrix elements of $M_\nu$ in Eq.~(20) are complex. This also corresponds to the case in which no generalized CP symmetry is imposed on the Yukawa Lagrangian. With this consideration, $M_\nu$ becomes complex symmetric and can be diagonalized by the unitary matrix given in Eq.~(24), but with an additional phase $\psi$;

\begin{equation}
M_\nu^{\text{diag}} = U_\nu^\dagger M_\nu U_\nu^*
\end{equation}
where
\begin{equation}
U_\nu =
\begin{pmatrix}
\cos\theta & 0 & \sin\theta\, e^{-i\psi} \\
0 & 1 & 0 \\
-\sin\theta\, e^{i\psi} & 0 & \cos\theta
\end{pmatrix}.
\end{equation}
With the above neutrino mixing matrix and the charged-lepton mass
diagonalizing matrix given in Eq.~(14), the lepton mixing matrix becomes
\begin{equation}
U = U_\ell^\dagger U_\nu = \frac{1}{\sqrt{3}}
\begin{pmatrix}
\cos\theta + \sin\theta\, e^{-i\psi} & 1 & \cos\theta - \sin\theta\, e^{i\psi} \\
\cos\theta + \omega \sin\theta\, e^{-i\psi}  & \omega^2 & \omega \cos\theta - \sin\theta\, e^{i\psi} \\
\cos\theta + \omega^2 \sin\theta\, e^{-i\psi} & \omega & \omega^2 \cos\theta - \sin\theta\, e^{i\psi}
\end{pmatrix}.
\end{equation}
Then the predictions for the lepton mixing angles are modified to
\begin{equation}
\sin^{2}\theta_{13}
=
\frac{1 - \sin 2\theta \cos\psi}{3},
\label{eq:sin2theta13}
\end{equation}

\begin{equation}
\sin^{2}\theta_{12}
=
\frac{1}{2 + \sin 2\theta \cos\psi},
\label{eq:sin2theta12}
\end{equation}

\begin{equation}
\sin^{2}\theta_{23}
=
\frac{
1 + \cos\theta \sin\theta
\left(
\cos\psi - \sqrt{3}\sin\psi
\right)
}{
2 + \sin 2\theta \cos\psi
}.
\label{eq:sin2theta23}
\end{equation}
Eq.~(43) reflects the non maximality of $\theta_{23}$ with respect to the maximal prediction in Eq.~(27). It is easy to see that for $\psi = 0$ it immediately reproduces the maximal prediction of $\theta_{23}$. It is important to note that the presence of the phase $\psi$ does not affect the expression of the Jarlskog invariant. Thus, the Jarlskog invariant corresponding to $U$ in Eq.~(40) is still given by Eq.~(30). The prediction of the non-maximal $\delta$ can be obtained by equating Eqs.~(29) and (30) followed by the substitution of the expressions of the lepton mixing angles from Eqs.~(41)--(43). The prediction of the Dirac CP phase $\delta$ turns out to be
\begin{equation}
\sin\delta
=
\pm
\left(
  1 +
  \frac{4 c^2 s^2 \sin^2\psi}
       {(c^2 - s^2)^2}
\right)^{-\tfrac{1}{2}}
\left(
  1 -
  \frac{3 c^2 s^2 \sin^2\psi}
       {(1 + c s \cos\psi)^2}
\right)^{-\tfrac{1}{2}},
\end{equation}
where the $'+'$ and $'-'$ sign corresponds to $\cos 2\theta > 0$ and $\cos 2\theta < 0$ respectively.

Having established the modified predictions for the lepton mixing angles and the Dirac CP phase resulting from the complex structure of the neutrino mass matrix, we now turn to the neutrino mass spectrum. The unitary diagonalization of the complex symmetric mass matrix in Eq.~(20) not only determines the mixing parameters but also fixes the light neutrino mass eigenvalues. These mass eigenvalues follow directly from the diagonalization of $M_\nu$ by the matrix $U_\nu$ given in Eq.~(39). Based on the above diagonalization, the corresponding neutrino mass
eigenvalues are given by
\begin{equation}
\begin{aligned}
m_1 &=  M_\nu^{11}\cos^2\theta
      + M_\nu^{13}\sin{2\theta}\,  e^{i\psi}
      + M_\nu^{33}\sin^2\theta\, e^{2i\psi}, \\
m_2 &= M_\nu^{22}, \\
m_3 &= M_\nu^{33}\cos^2\theta
      -  M_\nu^{13}\sin{2\theta}\, e^{-i\psi}
      + M_\nu^{11}\sin^2\theta\, e^{-2i\psi}.
\end{aligned}
\end{equation}
Since the neutrino mass eigenvalues obtained above are, in general,
complex, the physically relevant quantities entering neutrino
oscillation observables are their absolute squares. We therefore
consider the squared moduli of the mass eigenvalues, which can be
expressed in terms of the mass matrix elements, the mixing angle $\theta$, and the phase $\psi$. The resulting expressions for $|m_i|^2$ can be expressed as

\begin{equation}
\begin{aligned}
|m_1|^2
&= |M_{\nu}^{11}|^2
\Bigg[
\Big(
\cos^2\theta
+ \lambda_1 e^{i(\phi_{13}-\phi_{11})}\sin 2\theta \cos\psi
+ \lambda_2 e^{i(\phi_{33}-\phi_{11})}\sin^2\theta \cos 2\psi
\Big)^2 \\
&\hspace{1.0cm}
+
\Big(
\lambda_1 e^{i(\phi_{13}-\phi_{11})}\sin 2\theta \sin\psi
+ \lambda_2 e^{i(\phi_{33}-\phi_{11})}\sin^2\theta \sin 2\psi
\Big)^2
\Bigg],
\end{aligned}
\label{eq:m1sq}
\end{equation}

\begin{equation}
|m_2|^2 = |M_\nu^{22}|^2 ,
\label{eq:m2sq}
\end{equation}
\begin{equation}
\begin{aligned}
|m_3|^2
&=
|M_{\nu}^{11}|^2
\Bigg[
\Big(
\lambda_2 e^{i(\phi_{33}-\phi_{11})}\cos^2\theta
- \lambda_1 e^{i(\phi_{13}-\phi_{11})}\sin 2\theta \cos\psi \\
&\hspace{2.3cm}
+ \sin^2\theta \cos 2\psi
\Big)^2
+
\Big(
\lambda_1 e^{i(\phi_{13}-\phi_{11})}\sin 2\theta \sin\psi
- \sin^2\theta \sin 2\psi
\Big)^2
\Bigg].
\end{aligned}
\label{eq:m3sq}
\end{equation}
In the above expressions, we define the phases $\phi_{ij}$'s ($ij = 11,\,22,\,33,\,13$) throgh the polar forms $M_{\nu}^{ij} = \left| M_{\nu}^{ij} \right| e^{\, i \phi_{ij}}$. Further the parameters $\lambda_1$ and $\lambda_2$ represent ratios of absolute values of mass matrix elements:    
\begin{equation}
\lambda_1 = \frac{|M_\nu^{13}|}{|M_\nu^{11}|},
\qquad
\lambda_2 = \frac{|M_\nu^{33}|}{|M_\nu^{11}|}.
\end{equation}
To simplify the expressions we further consider specific choices of the relative phases $\phi_{13} - \phi_{11} = 0$ and
$\phi_{33} - \phi_{11} = 0$.
With these phase constraints and parameter definitions, we now proceed with a
systematic numerical analysis of the model.
We begin the numerical analysis by focusing on the model parameters
$\theta$ and $\psi$. Their allowed ranges are determined through a
correlation study based on Eqs.~(41) and (42). The parameters are
constrained such that the resulting values of $\sin\theta_{13}$ and
$\sin\theta_{12}$ fall within the experimentally allowed ranges reported
by the global oscillation analysis summarized in Table~1.

\FloatBarrier
\begin{figure}[H]
    \centering
    \includegraphics[width=0.60\textwidth]{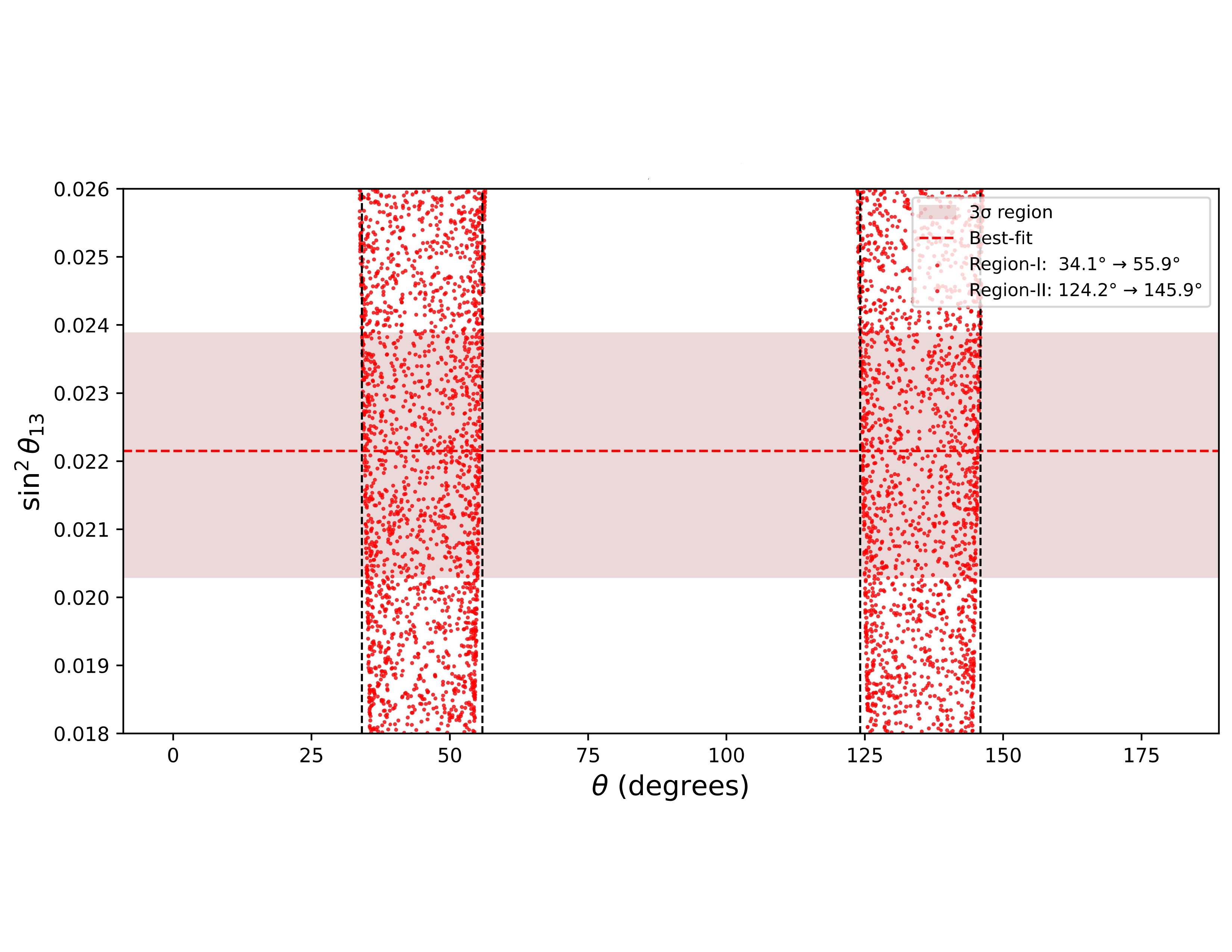}
    \caption{
    Correlation between $\sin^2\theta_{13}$ and the parameter $\theta$. The horizontal shaded band denotes the current $3\sigma$ allowed experimental range of $\sin^2\theta_{13}$, while the red dashed line indicates the best-fit value.
    }
    \label{fig:sin2theta13_theta}
\end{figure}

\begin{figure}[H]
    \centering
    \includegraphics[width=0.60\textwidth]{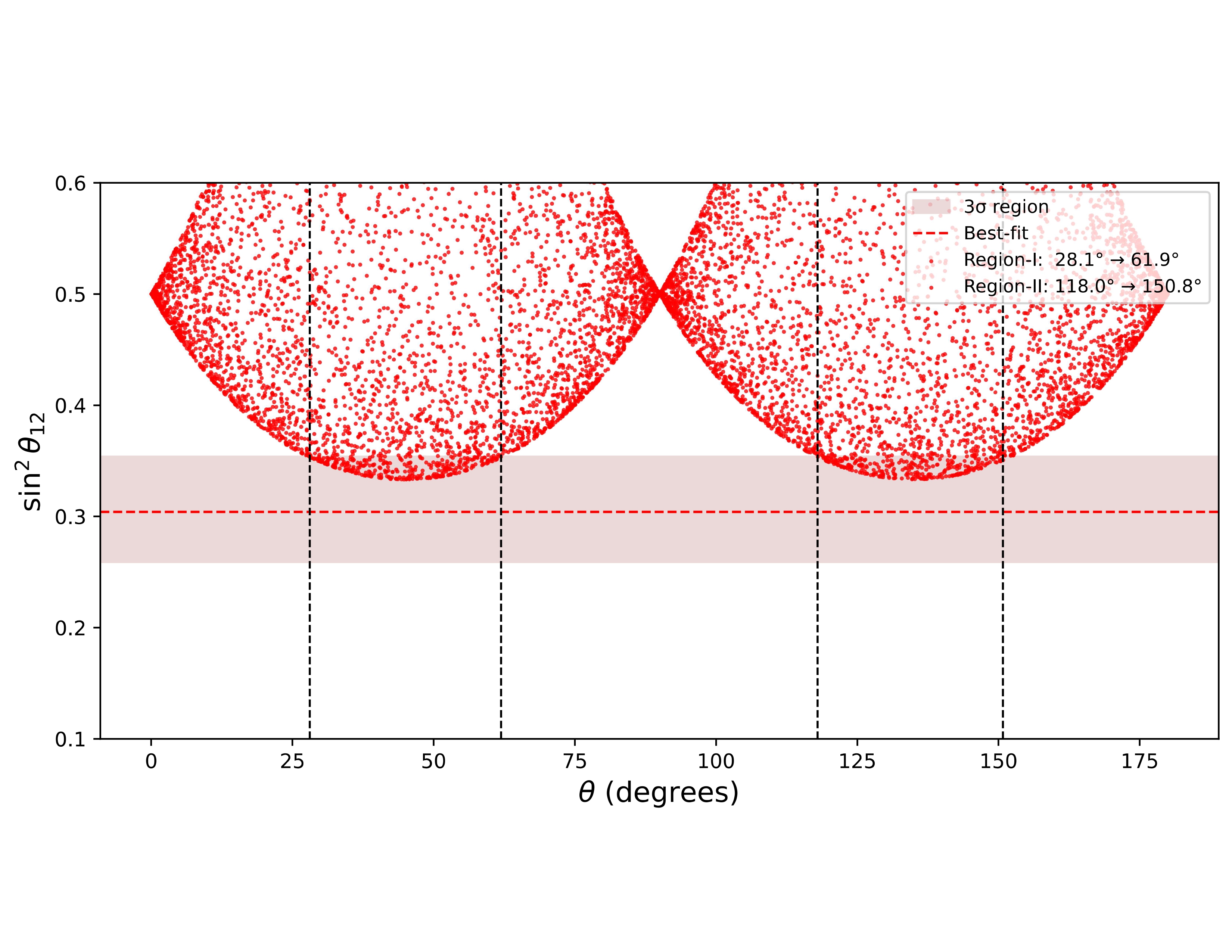}
    \caption{
    Correlation between $\sin^2\theta_{12}$ and the model parameter $\theta$. The horizontal shaded band denotes the current $3\sigma$ allowed experimental range of $\sin^2\theta_{12}$, while the red dashed line indicates the best-fit value.
    }
    \label{fig:sinsqtheta12_theta}
\end{figure}

\FloatBarrier

Figs.~\ref{fig:sin2theta13_theta} and \ref{fig:sinsqtheta12_theta}
illustrate the correlations of $\sin^2\theta_{13}$ and $\sin^2\theta_{12}$
with the model parameter $\theta$, respectively.
These analyses follow from the analytical relations given in
Eqs.~(41) and (42), where the phase parameter $\psi$ is varied
over the range $0^\circ \leq \psi \leq 360^\circ$.
Imposing the current $3\sigma$ experimental constraints on the
mixing angles, we find that the parameter $\theta$ is restricted
to two distinct allowed regions. For $\sin^2\theta_{13}$, the allowed regions are given by
\begin{equation}
\label{eq:theta13_allowed_regions}
34.1^\circ \leq \theta \leq 55.9^\circ
\quad, \quad
124.2^\circ \leq \theta \leq 145.9^\circ ,
\end{equation}
while for $\sin^2\theta_{12}$, they are
\begin{equation}
\label{eq:theta_allowed_regions}
28.1^\circ \leq \theta \leq 61.9^\circ,
\quad
118.0^\circ \leq \theta \leq 150.8^\circ.
\end{equation}
From Eqs.~\eqref{eq:theta13_allowed_regions} and \eqref{eq:theta_allowed_regions},
it is evident that the allowed ranges of $\theta$ obtained from the
$\sin^2\theta_{13}$ correlation are fully contained within those derived
from $\sin^2\theta_{12}$.
Hence, in the following analysis, we confine our discussion to the
$\theta$ intervals obtained from the $\sin^2\theta_{13}$ relation
given in \eqref{eq:theta13_allowed_regions}.

\FloatBarrier

\begin{figure}[H]
    \centering
    \includegraphics[width=0.60\textwidth]{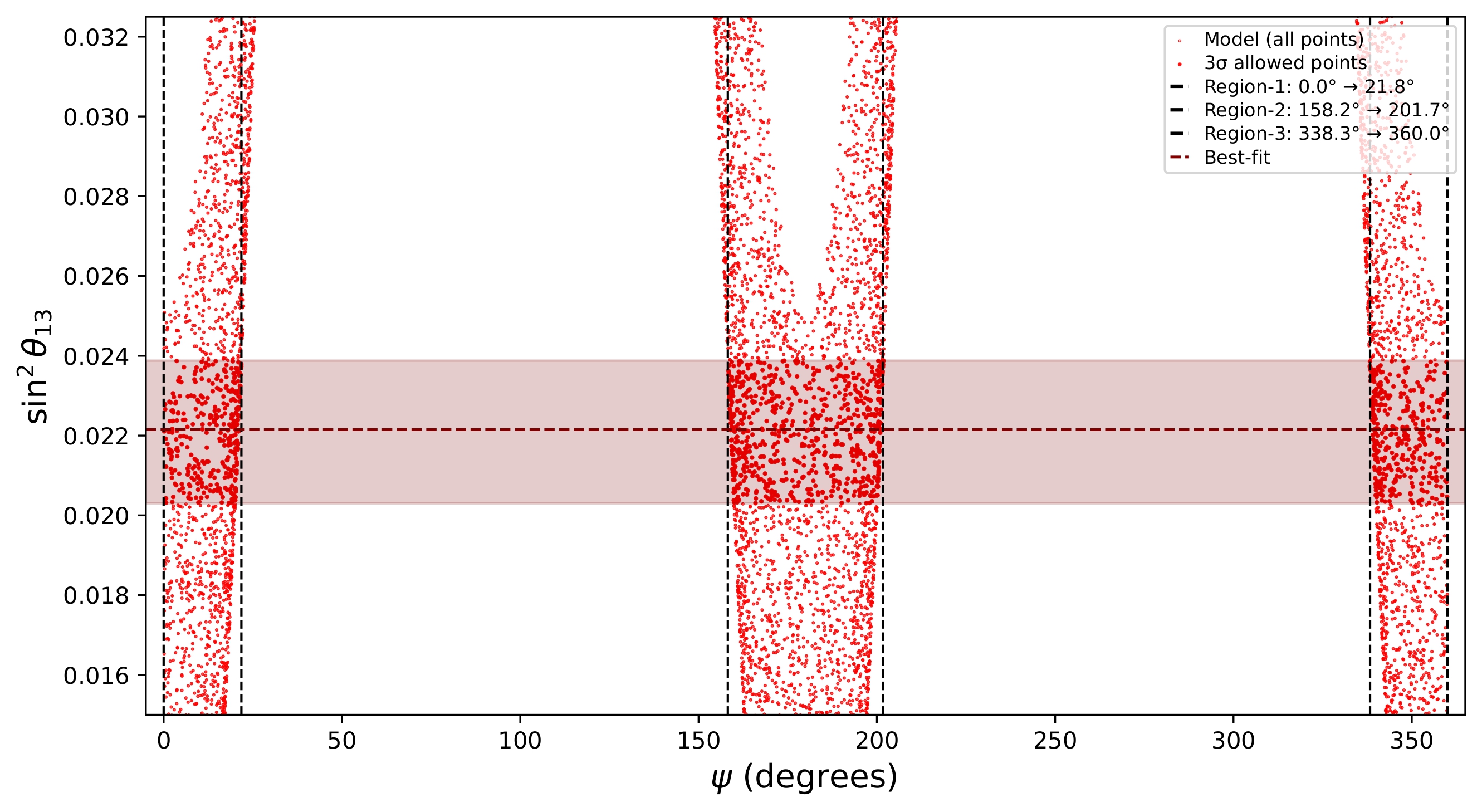}
    \caption{
    Correlation between $\sin^2\theta_{13}$ and the model parameter $\psi$.
    The horizontal shaded band denotes the current $3\sigma$ allowed experimental range of $\sin^2\theta_{13}$, while the red dashed line indicates the best-fit value.
    }
    \label{fig:theta13_vs_psi}
\end{figure}

\begin{figure}[H]
    \centering
    \includegraphics[width=0.60\textwidth]{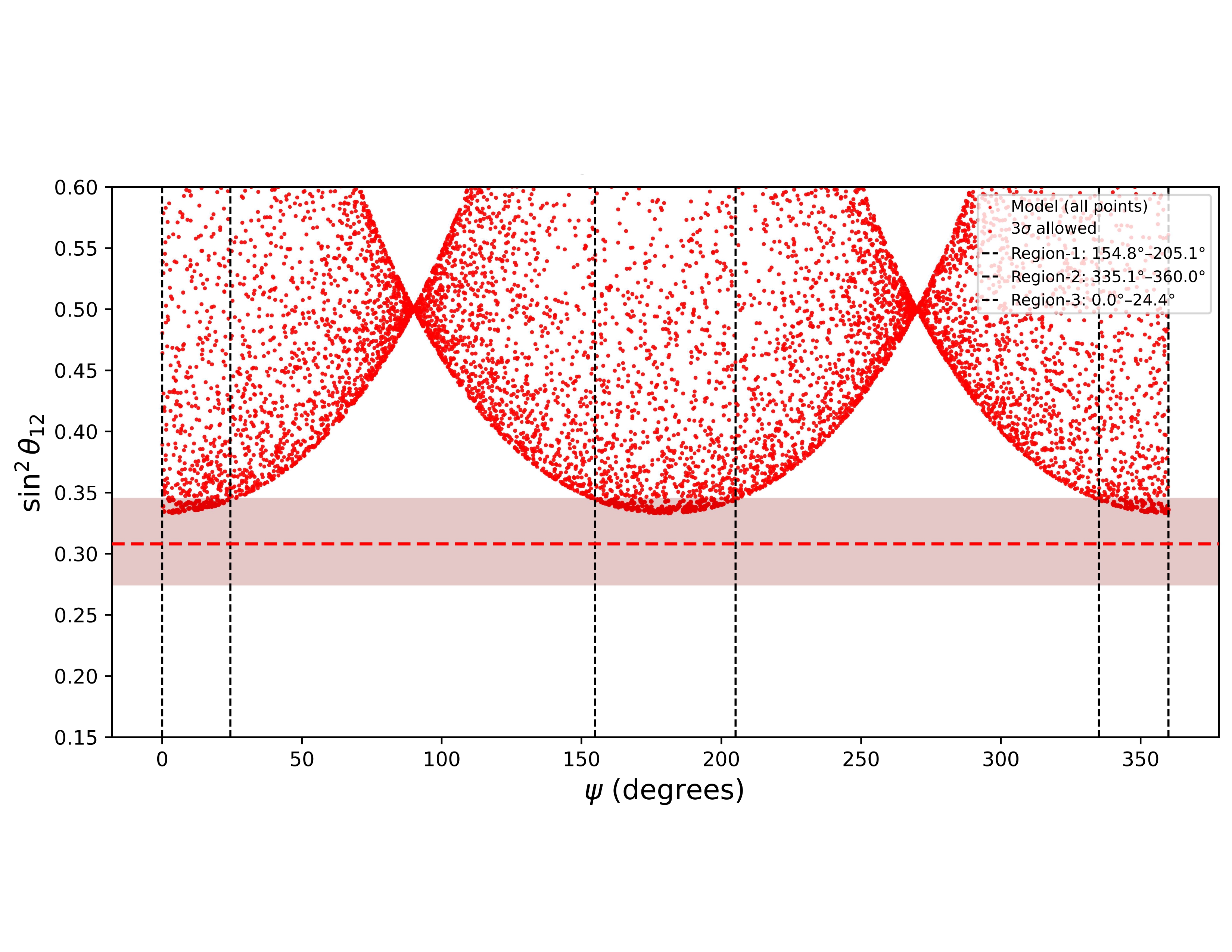}
    \caption{
    Correlation between $\sin^2\theta_{12}$ and the model parameter $\psi$.
    The horizontal shaded band denotes the current $3\sigma$ allowed experimental range of $\sin^2\theta_{12}$, while the red dashed line indicates the best-fit value.
    }
    \label{fig:theta12_vs_psi}
\end{figure}

\FloatBarrier

Figs.~\ref{fig:theta13_vs_psi} and \ref{fig:theta12_vs_psi}
display the correlations of $\sin^2\theta_{13}$ and $\sin^2\theta_{12}$
with the phase parameter $\psi$, respectively.
These correlation plots are obtained using the analytical expressions presented in
Eqs.~(41) and (42), with the parameter $\theta$ scanned over the range
$0^\circ \leq \theta \leq 180^\circ$.
Upon applying the current $3\sigma$ experimental bounds on the mixing angles,
the parameter $\psi$ is found to lie within three distinct allowed regions. For $\sin^2\theta_{13}$, the allowed regions are given by
\begin{equation}
\label{eq:psi_theta13_allowed}
0.0^\circ \leq \psi \leq 21.8^\circ
\;\; , \;\;
158.2^\circ \leq \psi \leq 201.7^\circ
\;\; , \;\;
338.3^\circ \leq \psi \leq 360.0^\circ ,
\end{equation}
while for $\sin^2\theta_{12}$, they are
\begin{equation}
\label{eq:psi_allowed_regions}
0.0^\circ \leq \psi \leq 24.4^\circ,\qquad
154.8^\circ \leq \psi \leq 205.1^\circ,\qquad
335.1^\circ \leq \psi \leq 360^\circ.
\end{equation}
Comparing Eqs.~\eqref{eq:psi_theta13_allowed} and
\eqref{eq:psi_allowed_regions}, it is evident that the allowed
regions of the phase parameter $\psi$ obtained from the
$\sin^2\theta_{13}$ correlation are entirely contained within
those derived from $\sin^2\theta_{12}$.
Accordingly, in the subsequent analysis, we focus on the
$\psi$ ranges permitted by the $\sin^2\theta_{13}$ relation.

\begin{figure}[H]
    \centering
    \includegraphics[width=0.60\textwidth]{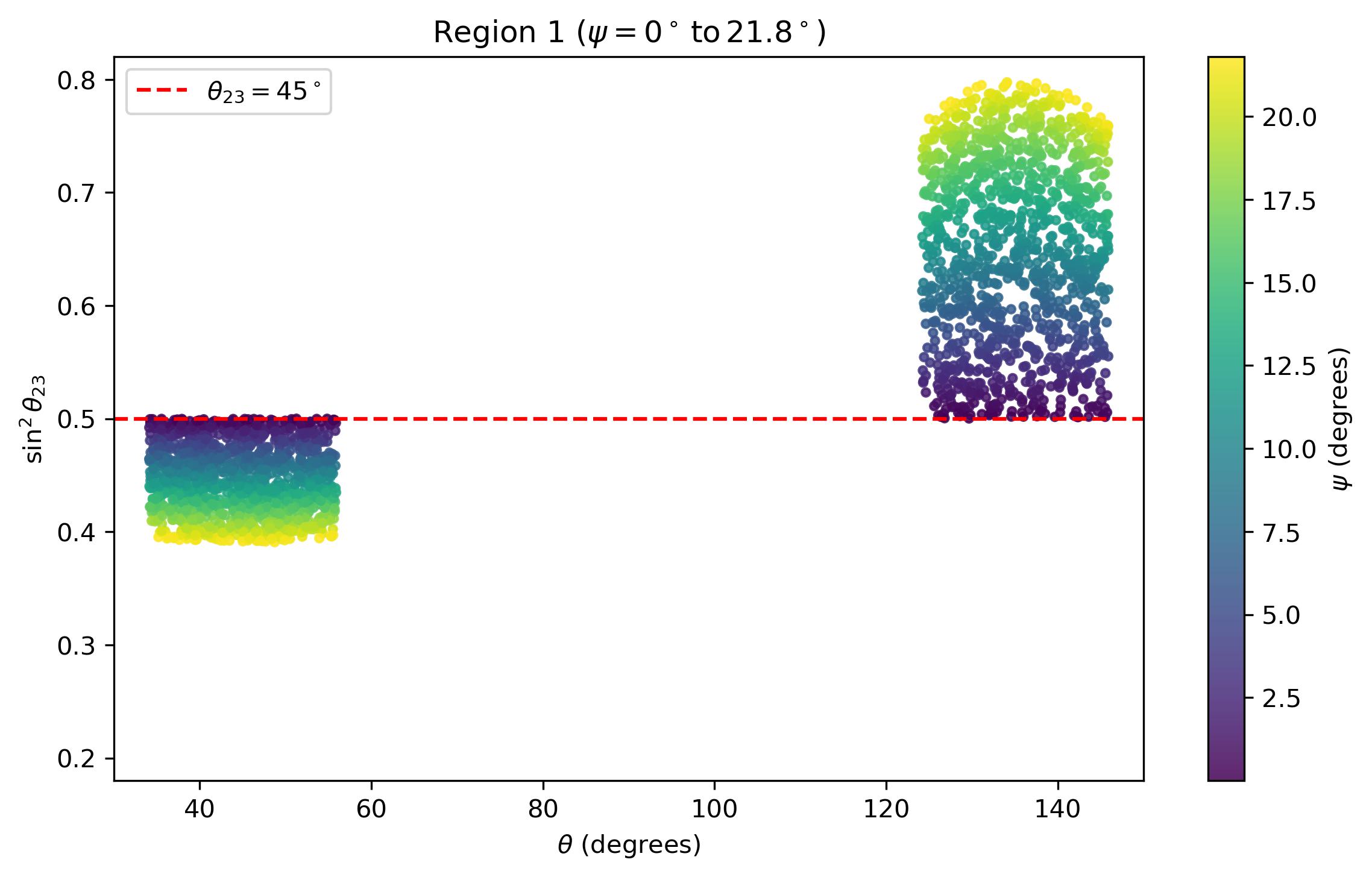}
    \caption{
        Correlation between $\sin^{2}\theta_{23}$ and $\theta$ for $0.0^\circ \leq \psi \leq 21.8^\circ$. The horizontal red dashed line represents the maximal value of $\sin^{2}\theta_{23}$. The color gradient along the vertical direction indicates that the deviation of $\sin^{2}\theta_{23}$ from its maximal value ($\sin^{2}\theta_{23} = 0.5$) increases with increasing $\psi$.
    }
    \label{fig:s23_theta_psi_0_21}
\end{figure}

Using the allowed ranges of the model parameters $\theta$ and $\psi$ obtained from the correlation analysis mentioned above, we next study the model 
predictions for $\sin^2\theta_{23}$ and $\delta$ using Eqs.~(43) and (44). We first study the correlation betwwen  $\sin^2\theta_{23}$ and $\theta$ using Eq.~(43). Since three distinct allowed regions of $\psi$ emerge from the previous analysis, we generate separate correlation plots of $\sin^2\theta_{23}$ versus $\theta$ for each permitted interval of $\psi$. The plots corresponding to the ranges $0.0^\circ \leq \psi \leq 21.8^\circ$, $158.2^\circ \leq \psi \leq 201.7^\circ$, and $338.3^\circ \leq \psi \leq 360.0^\circ$ are presented in Figs.~5, 6, and 7, respectively. In each plot, the color gradient along the vertical direction represents the variation of $\psi$. The horizontal red dashed line in each figure represents the maximal value $\sin^2\theta_{23} = 0.5$. From Fig.~\ref{fig:s23_theta_psi_0_21}, we observe that the range of the model parameter $\theta$ within $34.1^\circ \leq \theta \leq 55.9^\circ$, $\sin^2\theta_{23}$ allows $\sin^2\theta_{23}$ to lie in the first octant, whereas the range $124.2^\circ \leq \theta \leq 145.9^\circ$ leads to predictions in the second octant. The color gradient along the vertical direction indicates that the deviation of $\sin^2\theta_{23}$ from its maximal value ($\sin^2\theta_{23} = 0.5$) increases with increasing $\psi$ within the interval $0.0^\circ \leq \psi \leq 21.8^\circ$.

\begin{figure}[H]
    \centering
    \includegraphics[width=0.60\textwidth]{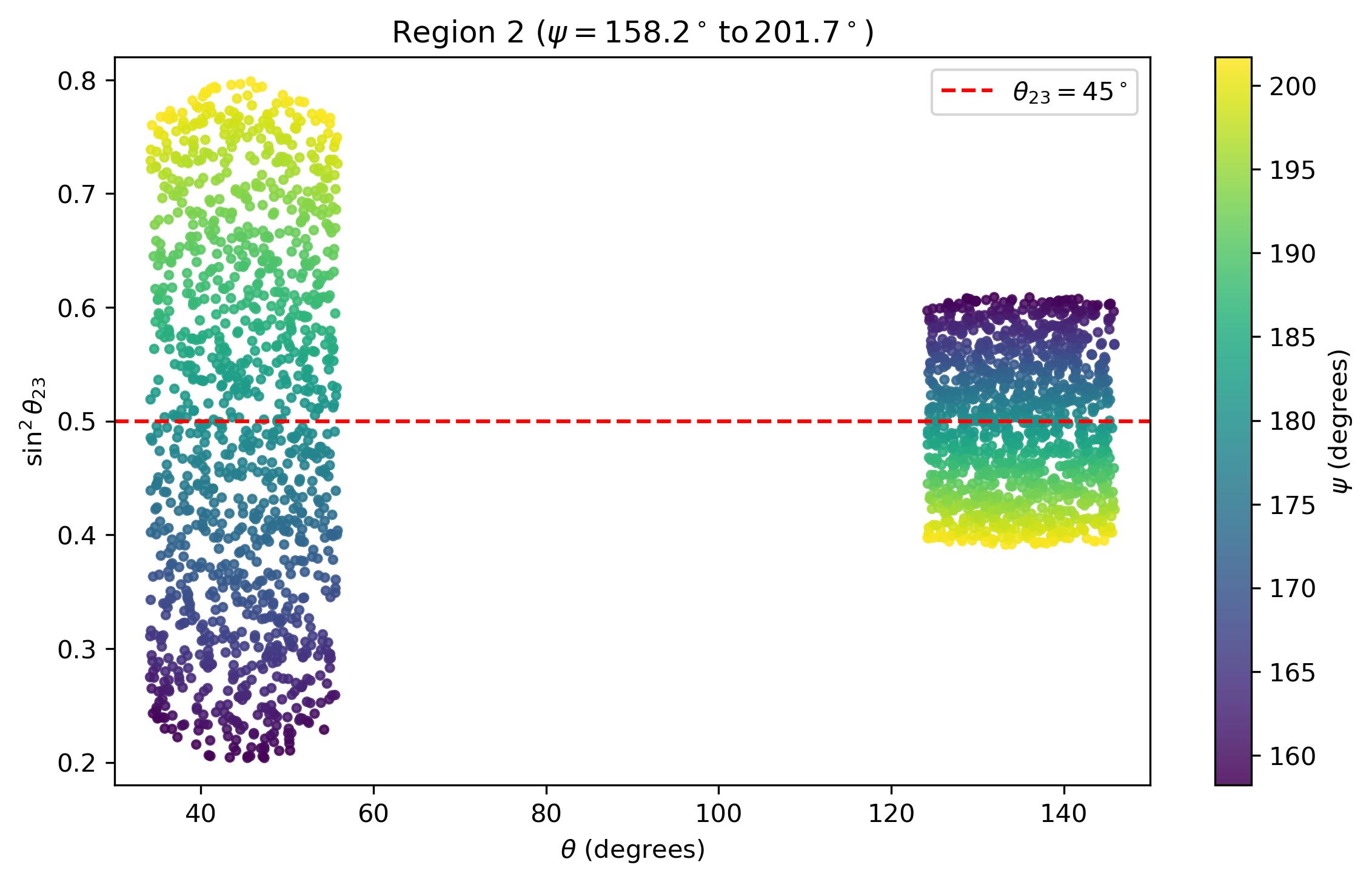}
    \caption{
        Correlation between $\sin^{2}\theta_{23}$ and $\theta$ for $158.2^\circ \leq \psi \leq 201.7^\circ$. The horizontal red dashed line represents the maximal value of $\sin^{2}\theta_{23}$. The color gradient along the vertical direction indicates that the deviation of $\sin^{2}\theta_{23}$ from its maximal value ($\sin^{2}\theta_{23} = 0.5$) increases with increasing $\psi$.
    }
    \label{fig:s23_theta_psi_158_201}
\end{figure}

Fig.~\ref{fig:s23_theta_psi_158_201} shows the correlation between $\sin^{2}\theta_{23}$ and $\theta$, corresponding to the two allowed $\theta$ ranges $34.1^\circ \leq \theta \leq 55.9^\circ$ and $124.2^\circ \leq \theta \leq 145.9^\circ$, for $158.2^\circ \leq \psi \leq 201.7^\circ$. It is evident from the figure that within this interval of $\psi$, $\sin^{2}\theta_{23}$ spans both the first and second octants for each of the allowed $\theta$ ranges. For the $\theta$ range $34.1^\circ \leq \theta \leq 55.9^\circ$, as $\psi$ decreases from approximately $180^\circ$, $\sin^2\theta_{23}$ deviates from its maximal value toward the first octant. In contrast, as $\psi$ increases from approximately $180^\circ$, $\sin^2\theta_{23}$ shifts toward the second octant. However, this behavior reverses in the $\theta$ range $124.2^\circ \leq \theta \leq 145.9^\circ$. In this case, decreasing $\psi$ from approximately $180^\circ$ drives $\sin^2\theta_{23}$ toward the second octant, while increasing $\psi$ shifts it towards the first octant.



\begin{figure}[H]
    \centering
    \includegraphics[width=0.60\textwidth]{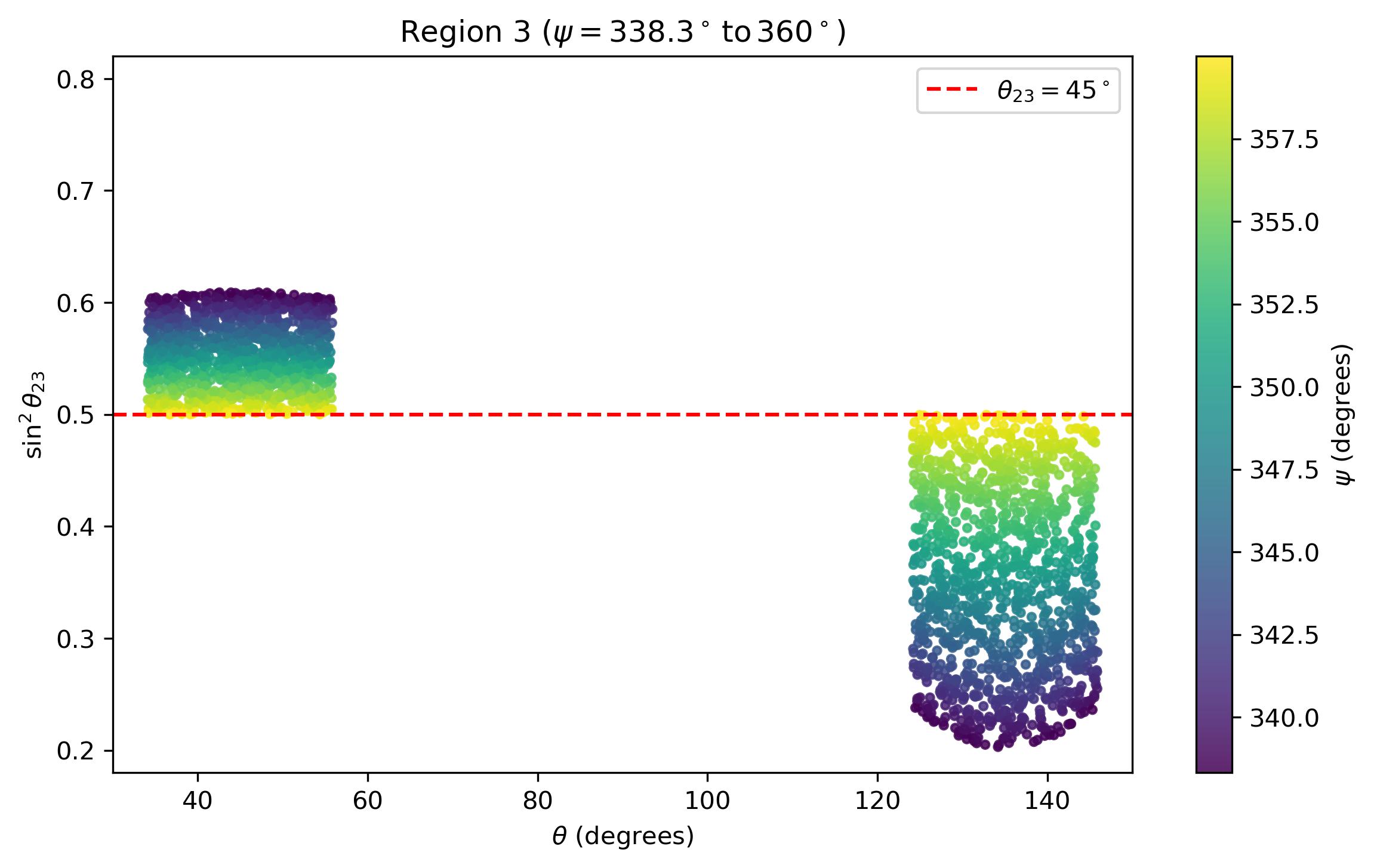}
    \caption{
        Correlation between $\sin^{2}\theta_{23}$ and $\theta$ for $338.3^\circ \leq \psi \leq 360^\circ$. The horizontal red dashed line represents the maximal value of $\sin^{2}\theta_{23}$. The color gradient along the vertical direction indicates that the deviation of $\sin^{2}\theta_{23}$ from its maximal value ($\sin^{2}\theta_{23} = 0.5$) increases with increasing $\psi$.
    }
    \label{fig:s23_theta_psi_338_359}
\end{figure}

Similarly, Fig.~\ref{fig:s23_theta_psi_338_359} illustrates the correlation between $\sin^2\theta_{23}$ and $\theta$ for $158.2^\circ \leq \psi \leq 201.7^\circ$. In this case, the behavior of $\sin^2\theta_{23}$ as a function of $\theta$ is reversed compared to that shown in Fig.~\ref{fig:s23_theta_psi_0_21}. In this case, the range $34.1^\circ \leq \theta \leq 55.9^\circ$, as $\psi$ of the model parameter $\theta$ yields $\sin^2\theta_{23}$ in the second octant, whereas $124.2^\circ \leq \theta \leq 145.9^\circ$, as $\psi$ corresponds to the first octant. The color gradient along the vertical direction indicates that the deviation of $\sin^2\theta_{23}$ from its maximal value ($\sin^2\theta_{23}=0.5$) increases with decreasing $\psi$ within the interval $338.3^\circ \leq \psi \leq 360^\circ$.
\FloatBarrier

\FloatBarrier
\FloatBarrier

\begin{figure}[H]
    \centering
    \includegraphics[width=0.60\textwidth]{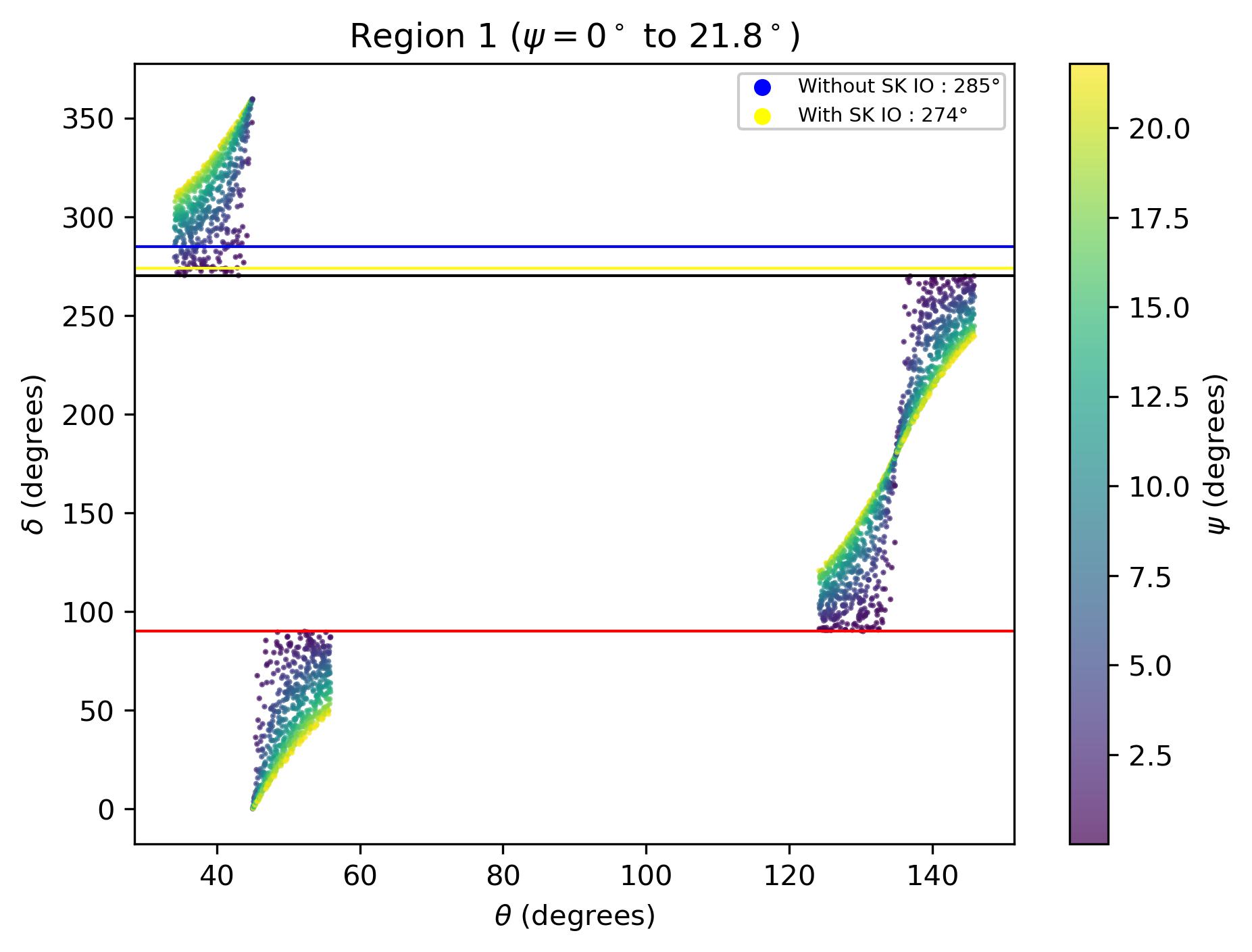}
    \caption{
        Correlation between $\delta$ and $\theta$ for $0.0^\circ \leq \psi \leq 21.8^\circ$. The black and red solid lines denote the maximal values of $\delta$, namely $270^\circ$ and $90^\circ$, respectively, while the blue and yellow lines indicate the best-fit values in the IO without and with SK data, respectively. The colour gradient along the vertical axis shows increasing $\psi$.
    }
    \label{fig:delta_theta_psi_0_21}
\end{figure}

\begin{figure}[H]
    \centering
    \includegraphics[width=0.60\textwidth]{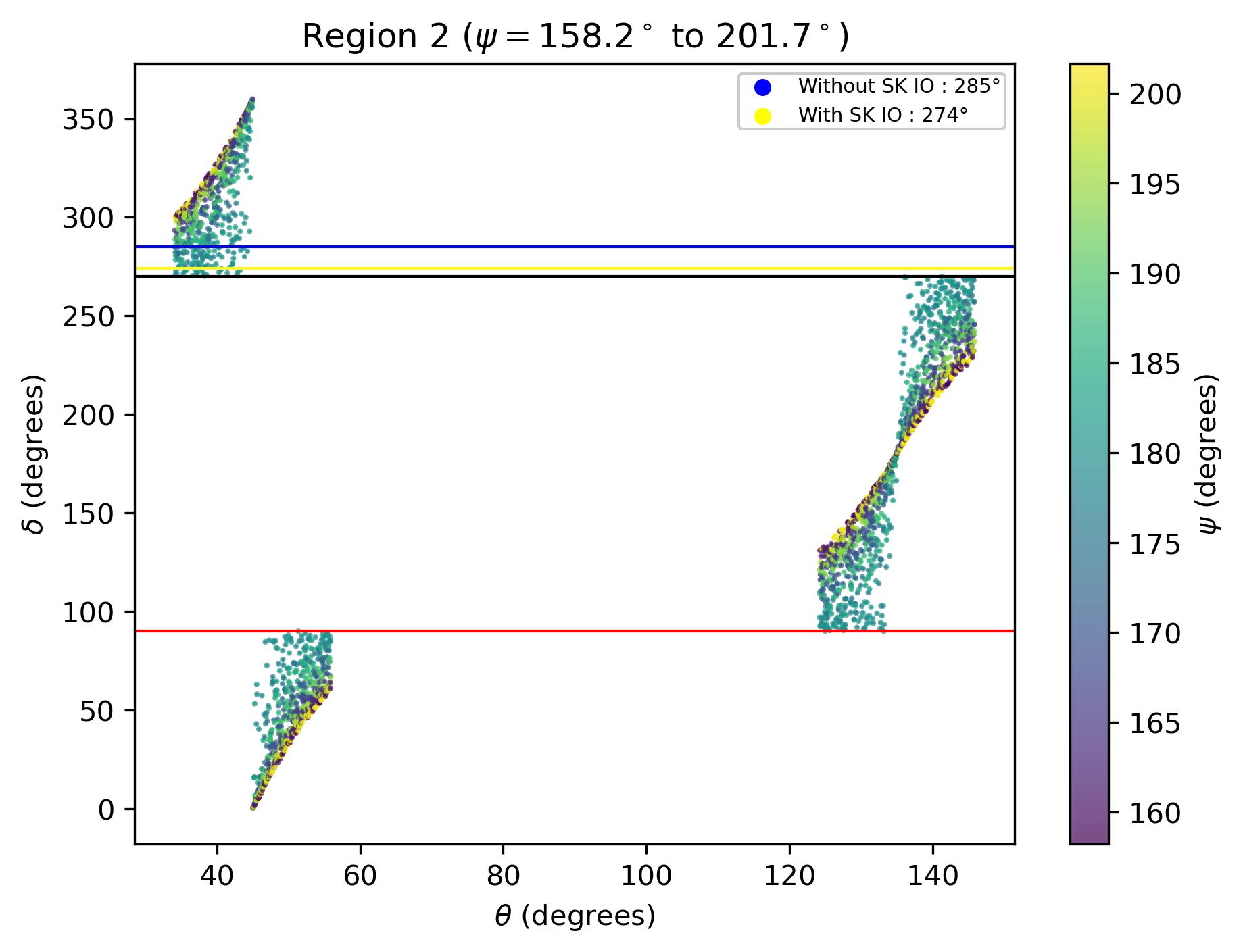}
    \caption{
       Correlation between $\delta$ and $\theta$ for $158.2^\circ \leq \psi \leq 201.7^\circ$. The black and red solid lines denote the maximal values of $\delta$, namely $270^\circ$ and $90^\circ$, respectively, while the blue and yellow lines indicate the best-fit values in the IO without and with SK data, respectively. The colour gradient along the vertical axis shows increasing $\psi$.
    }
    \label{fig:delta_theta_psi_158_201}
\end{figure}

\begin{figure}[H]
    \centering
    \includegraphics[width=0.60\textwidth]{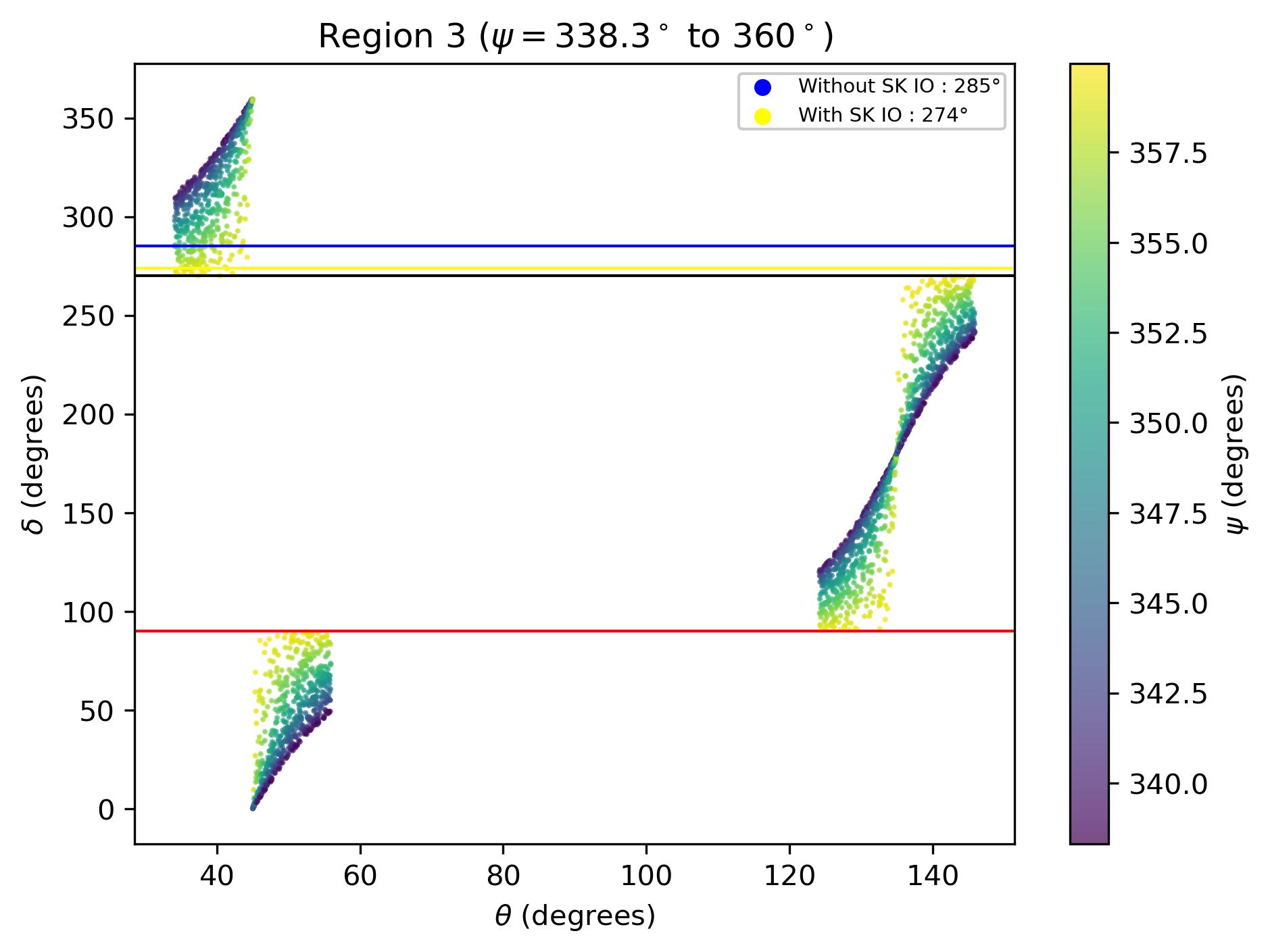}
    \caption{
   Correlation between $\delta$ and $\theta$ for $338.3^\circ \leq \psi \leq 360^\circ$. The black and red solid lines denote the maximal values of $\delta$, namely $270^\circ$ and $90^\circ$, respectively, while the blue and yellow lines indicate the best-fit values in the IO without and with SK data, respectively. The colour gradient along the vertical axis shows increasing $\psi$.
    }
    \label{fig:delta_theta_psi_338_359}
\end{figure}

\FloatBarrier



In a similar way, we study the prediction of $\delta$ as a function of $\theta$ using Eq.~(44). In each correlation plot, four solid horizontal lines with different colors are shown. The black and red solid lines correspond to the maximal values of $\delta$, namely $270^\circ$ and $90^\circ$, respectively. The blue and yellow solid lines represent the best-fit values of $\delta$ in the inverted ordering scenario, without SK data and with SK data, respectively.
To generate the correlation plot between $\delta$ and $\theta$, we first vary $\psi$ within the range $0.0^\circ \leq \psi \leq 21.8^\circ$. The corresponding plot is shown in Fig.~\ref{fig:delta_theta_psi_0_21}. From this figure, we observe that the model prediction of $\delta$ lies close to the maximal value $90^\circ$ for the parameter ranges $45^\circ \leq \theta \leq 55.9^\circ$ and $124.5^\circ \leq \theta \leq 135^\circ$. On the other hand, the model prediction of $\delta$ approaches the maximal value $270^\circ$ for $34.1^\circ \leq \theta \leq 45^\circ$ and $135^\circ \leq \theta \leq 145.9^\circ$. We also observe that within the range $0^\circ \leq \psi \leq 21.8^\circ$, the maximal values $\delta = 90^\circ$ or $270^\circ$ are approximately obtained for lower values of $\psi$. However, as $\psi$ increases within this range, the predicted values of $\delta$ gradually deviate from their maximal values.


\noindent
The second correlation plot between $\delta$ and $\theta$ is generated by varying $\psi$ within the range $158.2^\circ \leq \psi \leq 201.7^\circ$. The corresponding plot is shown in Fig.~\ref{fig:delta_theta_psi_158_201}. From this figure also, we observe that the model prediction of $\delta$ lies close to the maximal value $90^\circ$ for the parameter ranges $45^\circ \leq \theta \leq 55.9^\circ$ and $124.5^\circ \leq \theta \leq 135^\circ$. On the other hand, the model prediction of $\delta$ approaches the maximal value $270^\circ$ for $34.1^\circ \leq \theta \leq 45^\circ$ and $135^\circ \leq \theta \leq 145.9^\circ$. The distinction, however, originates from the behaviour with respect to $\psi$. In this range, the maximal values $\delta = 270^\circ$ and $90^\circ$ are realized when $\psi$ is approximately close to $180^\circ$.
It is also evident from this plot that larger deviations of $\delta$ from its maximal values occur for the extreme lower and upper values of $\psi$ within this range.



Fig.~\ref{fig:delta_theta_psi_338_359} displays the correlation between the CP-violating phase $\delta$ and the model parameter $\theta$ for the third allowed interval $338.3^\circ \leq \psi \leq 360^\circ$. This plot shows the same qualitative structure as the previous cases. The predicted values of $\delta$ cluster around the maximal limits $90^\circ$ and $270^\circ$ within the respective $\theta$ intervals, demonstrating that the preference for maximal CP violation persists in this $\psi$ range as well. However, the distinction arises from the behavior of $\delta$ with respect to $\psi$ in this specific range, where the maximal values $\delta = 90^\circ$ and $270^\circ$ are approximately realized for higher values of $\psi$, i.e., when $\psi$ approaches $360^\circ$. However, as $\psi$ decreases within this interval, the predicted values of $\delta$ gradually deviate from their maximal values.

We now present a systematic and comprehensive analysis of the different allowed ranges of $\theta$ and $\psi$ in connection with the model predictions for the mixing angles and the Dirac CP phase, in agreement with the global data. We proceed separately for normal order (NO) and inverted order (IO) in the following two subsections.

\subsection{Normal Order (NO) scenario}

As per the global analysis of three-neutrino oscillation data (Table~1), we observe that that in the normal order (NO) scenario, the best-fit value of $\sin^{2}\theta_{23}$ lies in the second octant when the Super-Kamiokande (SK) atmospheric data are not included. However, once the SK data are incorporated, the preferred value of $\sin^{2}\theta_{23}$ shifts to the first octant. In view of this, we perform an analysis of the allowed regions of the model parameters $\theta$ and $\psi$ that can accommodate these experimental observations. We first search for parameter values that yield $\sin^{2}\theta_{23}$ in the second octant together with a Dirac CP phase $\delta$ close to its best-fit value $\delta \simeq 177^\circ$, corresponding to the global fit without SK data. 
From Fig.~\ref{fig:s23_theta_psi_0_21}, it is observed that
$\sin^{2}\theta_{23}$ lies in the second octant for the
parameter region $124.2^\circ \le \theta \le 145.9^\circ$
with the corresponding range $0^\circ \le \psi \le 21^\circ$.
However, to obtain only a small deviation from the maximal
value of $\sin^{2}\theta_{23}$, the allowed range of $\psi$
is further restricted to $0^\circ \le \psi \le 5^\circ$. On the other hand, Fig.~\ref{fig:delta_theta_psi_0_21} shows
that values of $\delta$ below $270^\circ$ are obtained for
$124.2^\circ \le \theta \le 145.9^\circ$ and
$0^\circ \le \psi \le 21.8^\circ$. In the parameter region
$124.2^\circ \le \theta \le 145.9^\circ$ and
$0^\circ \le \psi \le 5^\circ$, the predicted value of the
reactor mixing angle satisfies
$\sin^{2}\theta_{13} \gtrsim 0.64$, which is far above the
experimentally allowed $3\sigma$ range. Therefore, this
region of the parameter space is not found to be consistent with the desired experimental predictions. Turning to Fig.6, it is observed that
$\sin^{2}\theta_{23}$ lies in the second octant for both allowed
$\theta$ ranges. In particular, the second-octant solution is
obtained for $34.1^\circ \le \theta \le 55.9^\circ$ with
$180^\circ \lesssim \psi \lesssim 201.7^\circ$, as well as for
$124.2^\circ \le \theta \le 145.9^\circ$ with the corresponding
range $158.2^\circ \lesssim \psi \lesssim 180^\circ$. From Fig.~\ref{fig:delta_theta_psi_158_201}, it is further observed
that values of $\delta$ below $270^\circ$ are realized for
$124.2^\circ \le \theta \le 145.9^\circ$ and
$158.2^\circ \le \psi \le 201.7^\circ$. Therefore, the common parameter region that simultaneously yields
$\sin^{2}\theta_{23}$ in the second octant and $\delta < 270^\circ$
is given by
$124.2^\circ \le \theta \le 145.9^\circ$ and
$158.2^\circ \le \psi \le 180^\circ$. Within these suitable parameter ranges, we also look for specific values of the model parameters $\theta$ and $\psi$ that yield mixing angles and the Dirac CP phase in best agreement with the global data. As an example, we choose
$\theta = 145.2^\circ$ and $\psi = 179^\circ$,
for which we obtain
$\sin^{2}\theta_{13} \approx 0.0212$, in good agreement with
the best-fit value $0.02195$.
The corresponding prediction for the atmospheric mixing angle
is $\sin^{2}\theta_{23} \approx 0.505$, which lies in the
second octant and remains reasonably close to its best-fit
value $0.561$.
Furthermore, the Dirac CP phase is found to be
$\delta \approx 170^\circ$, which is consistent with best-fit value
$\delta \simeq 177^\circ$.

Fig.~\ref{fig:s23_theta_psi_338_359} shows that
$\sin^{2}\theta_{23}$ lies in the second octant for
$34.1^\circ \le \theta \le 55.9^\circ$ with the corresponding
range $338.3^\circ \le \psi \le 360^\circ$. In contrast, Fig.~\ref{fig:delta_theta_psi_338_359}
indicates that values of $\delta$ below $270^\circ$
occur for $124.2^\circ \le \theta \le 145.9^\circ$
within the same $\psi$ interval. Since there is no overlapping $\theta$ range that
simultaneously yields $\sin^{2}\theta_{23}$ in the
second octant and $\delta < 270^\circ$ for
$338.3^\circ \le \psi \le 360^\circ$,
this region of the $(\theta,\psi)$ parameter space
is inconsistent with the experimental observations.

\begin{figure}[!t]
    \centering
    \includegraphics[width=0.60\textwidth]{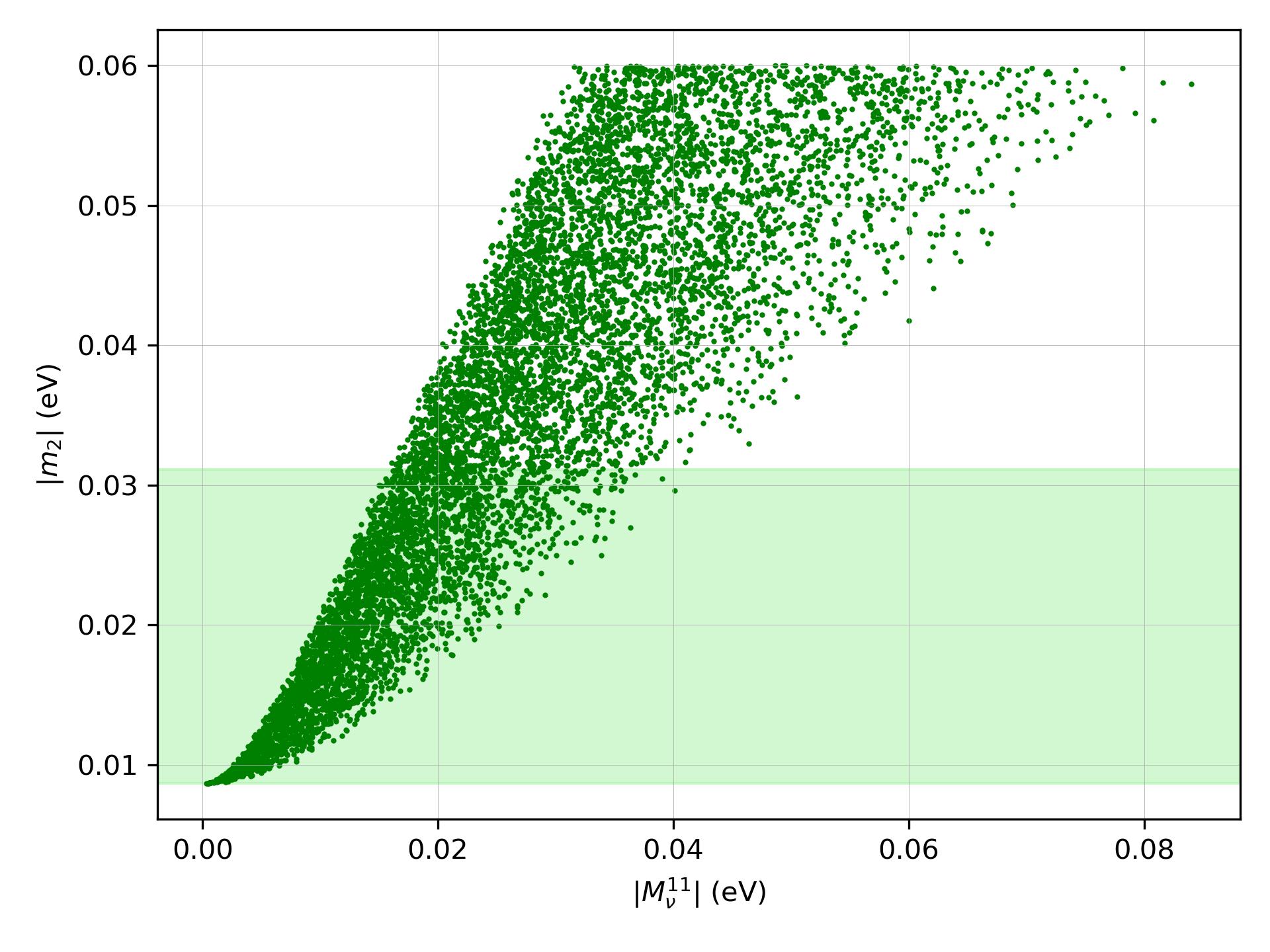}
    \caption{Correlation between the mass eigenvalue $|m_2|$ and the neutrino 
mass matrix element $|(M_{\nu})_{11}|$ for the normal mass ordering (NO).}
    \label{fig:AC_correlation}
\end{figure}

We now proceed to determine allowed parameter regions that can predict $\sin^{2}\theta_{23}$ in the first octant while simultaneously predicting $\delta$ close to $\delta \simeq 212^\circ$, as preferred when the SK data are included. From the correlation plot of $\sin^{2}\theta_{23}$ versus $\theta$
shown in Fig.~\ref{fig:s23_theta_psi_0_21}, it is evident that
$\sin^{2}\theta_{23}$ lies in the first octant for the range $34.1^\circ \leq \theta \leq 55.9^\circ$,
with the corresponding allowed range
$0^\circ \leq \psi \leq 21.8^\circ$.
However, from Fig.~\ref{fig:delta_theta_psi_0_21}, which shows the
variation of the Dirac CP phase $\delta$ with $\theta$ in the same
$\psi$ interval, it is observed that values of $\delta$ close to $\delta \simeq 212^\circ$ is obtained only for
$124.2^\circ \leq \theta \leq 145.9^\circ$, $0^\circ \leq \psi \leq 21.8^\circ$.
Since there is no common region in the $(\theta,\psi)$ parameter space
that simultaneously yields $\sin^{2}\theta_{23}$ in the first octant
and $\delta$ close to $\delta \simeq 212^\circ$, this region is excluded from further
consideration in our analysis. From Fig.~\ref{fig:s23_theta_psi_158_201}, it is observed that 
$\sin^{2}\theta_{23}$ lies in the first octant for two distinct 
$(\theta,\psi)$ regions within the range 
$158.2^\circ \leq \psi \leq 201.7^\circ$. 
The first region corresponds to
$34.1^\circ \leq \theta \leq 55.9^\circ$,
$160^\circ \lesssim \psi \lesssim 180^\circ$,
while the second region corresponds to
$124.2^\circ \leq \theta \leq 145.9^\circ$,
$180^\circ \lesssim \psi \lesssim 200^\circ$.
However, from Fig.~\ref{fig:delta_theta_psi_158_201}, it is seen that 
for $158.2^\circ \leq \psi \leq 201.7^\circ$, values of the Dirac CP phase 
$\delta$ close to $\delta \simeq 212^\circ$ are obtained only for
$135^\circ \leq \theta \leq 145.9^\circ$.
Therefore, the allowed parameter space is restricted to the 
overlapping portion of the second region, namely
$135^\circ \leq \theta \leq 145.9^\circ$,
$180^\circ \lesssim \psi \lesssim 200^\circ$,
which simultaneously yields the first-octant solution of 
$\theta_{23}$ and $\delta$ close to $\delta \simeq 212^\circ$. Within this allowed parameter space, we choose  
$\theta \approx 145.3^\circ$ and $\psi \approx 181^\circ$. 
For these values, the model predicts 
$\sin^{2}\theta_{13} \approx 0.0214$, 
which is very close to the best-fit value 
$\sin^{2}\theta_{13} = 0.02215$ for NO scenario including SK data. The corresponding prediction for the atmospheric mixing angle is 
$\sin^{2}\theta_{23} \approx 0.494$, 
which lies in the first octant and is reasonably close to its best-fit value $0.470$. Furthermore, the predicted Dirac CP phase is 
$\delta \approx 226^\circ$, 
which is fairly close to the global best-fit value $212^\circ$. For the range $338.3^\circ \leq \psi \leq 360^\circ$, 
Fig.~\ref{fig:s23_theta_psi_338_359} shows that 
$\sin^{2}\theta_{23}$ lies in the first octant for 
$124.2^\circ \leq \theta \leq 145.9^\circ$.
However, from Fig.~\ref{fig:delta_theta_psi_338_359}, 
values of the Dirac CP phase $\delta$ close to $\delta \simeq 212^\circ$ 
is obtained only in the more restricted range 
$135^\circ \leq \theta \leq 145.9^\circ$.
Thus, the overlapping region that simultaneously satisfies 
the first-octant solution for $\sin^{2}\theta_{23}$ and $\delta$ close to $\delta \simeq 212^\circ$  is 
$135^\circ \leq \theta \leq 145.9^\circ$, 
$338.3^\circ \leq \psi \leq 360^\circ$.
But, within this overlapping parameter space, the minimum 
predicted value of $\sin^{2}\theta_{13}$ is approximately $0.62$, 
which is significantly larger than its best-fit value 
$\sin^{2}\theta_{13} = 0.02215$. Therefore, this region of the 
parameter space is phenomenologically inconsistent and hence 
excluded from further analysis.

On the basis of the above analysis of the allowed ranges of $\theta$ and $\psi$, we perform a correlation analysis using Eqs.~(46)--(48) to study the relationship among the mass matrix elements $|{M_\nu}^{11}|$, $|{M_\nu}^{22}|$, and the parameters $\lambda_1$ and $\lambda_2$. For this purpose, we adopt the specific values of $\theta$ and $\psi$ identified in the previous discussion, which yield predictions for the mixing angles and the Dirac CP phase $\delta$ consistent with the global fit data. In addition, we use the best-fit values of the mass-squared differences $\Delta m_{21}^{2}$ and $\Delta m_{32}^{2}$, along with the cosmological 
upper bound on the sum of neutrino masses $\sum m_i$.

We note that the mass-squared differences 
$\Delta m_{21}^{2} = m_{2}^{2} - m_{1}^{2}$ and 
$|\Delta m_{32}^{2}| = |m_{3}^{2} - m_{2}^{2}|$ 
leave only one neutrino mass eigenvalue undetermined. 
Since $|m_{2}|$ is directly related to the matrix element 
$|(M_\nu)_{22}|$ (see Eq.~(47)), we treat $|m_{2}|$ as a free 
parameter in our correlation analysis. Using the experimental best-fit values of $\Delta m_{21}^{2}$ and $\Delta m_{32}^{2}$, together with the cosmological upper bound on the sum of neutrino masses 
$\sum_i m_i < 0.12\,\text{eV}$ \cite{RoyChoudhury2020}, we determine the corresponding 
allowed lower and upper bounds on $|m_2|$. For NO scenario, using the best-fit values $\Delta m_{21}^{2} = 7.49 \times 10^{-5}\text{eV}^{2}$ and $\Delta m_{32}^{2} = 2.513 \times 10^{-3}\text{eV}^{2}$, including SK data, we obtain
\begin{equation}
|m_2|^{\min} \approx 0.008654~\text{eV}, 
\qquad 
|m_2|^{\max} \approx 0.031113~\text{eV}.
\end{equation}
Notably, the minimum and maximum values of $|m_{2}|$ show negligible variation when SK data are excluded; therefore, we do not consider the case without SK data separately.

\begin{figure}[t]
    \centering
    \includegraphics[width=0.60\textwidth]{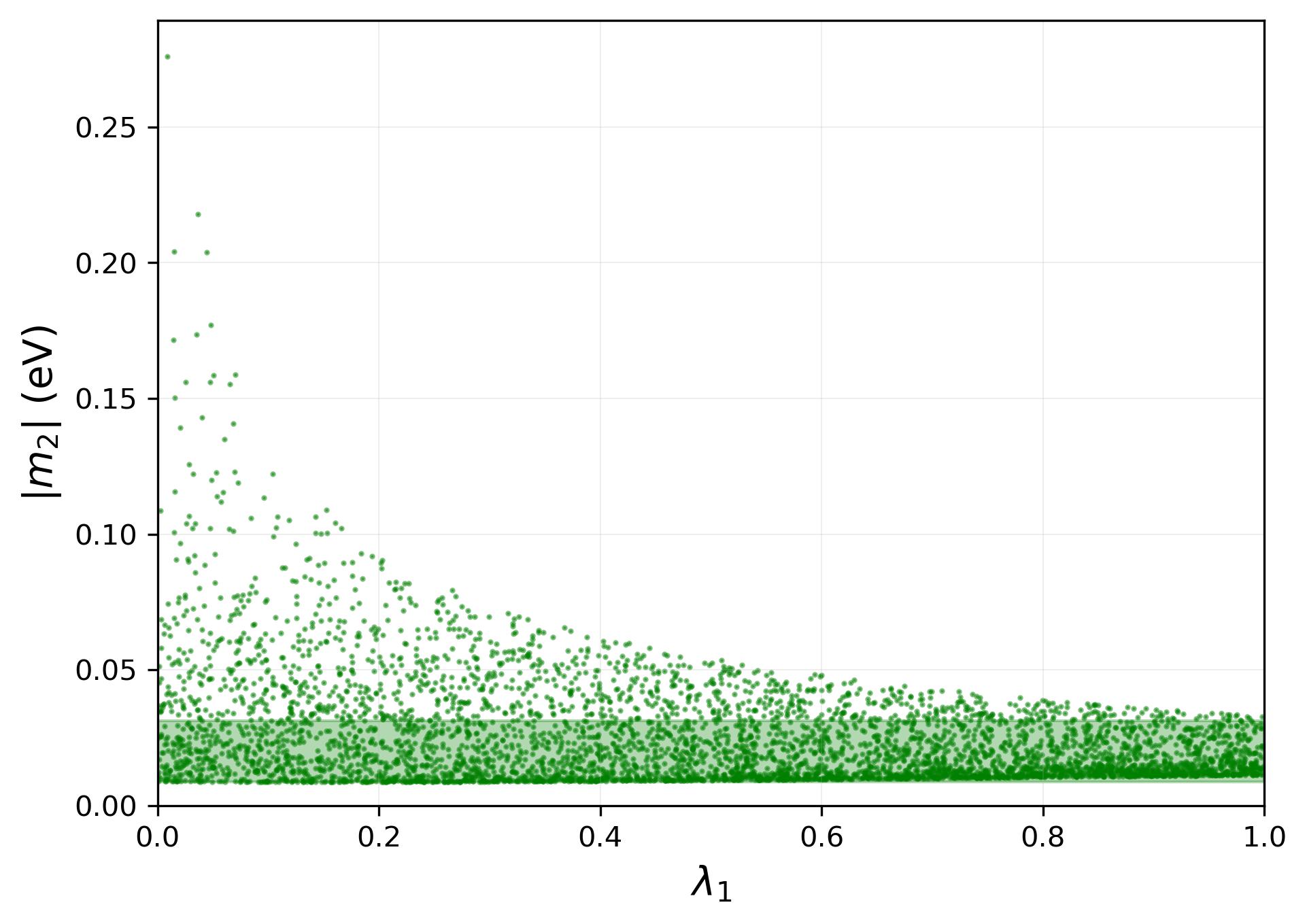}
    \caption{Correlation between the mass eigenvalue $|m_2|$ and $\lambda_1$ for the normal mass ordering (NO).}
    \label{fig:m2_vs_lambda1}
\end{figure}


\begin{figure}[t]
    \centering
    \includegraphics[width=0.60\textwidth]{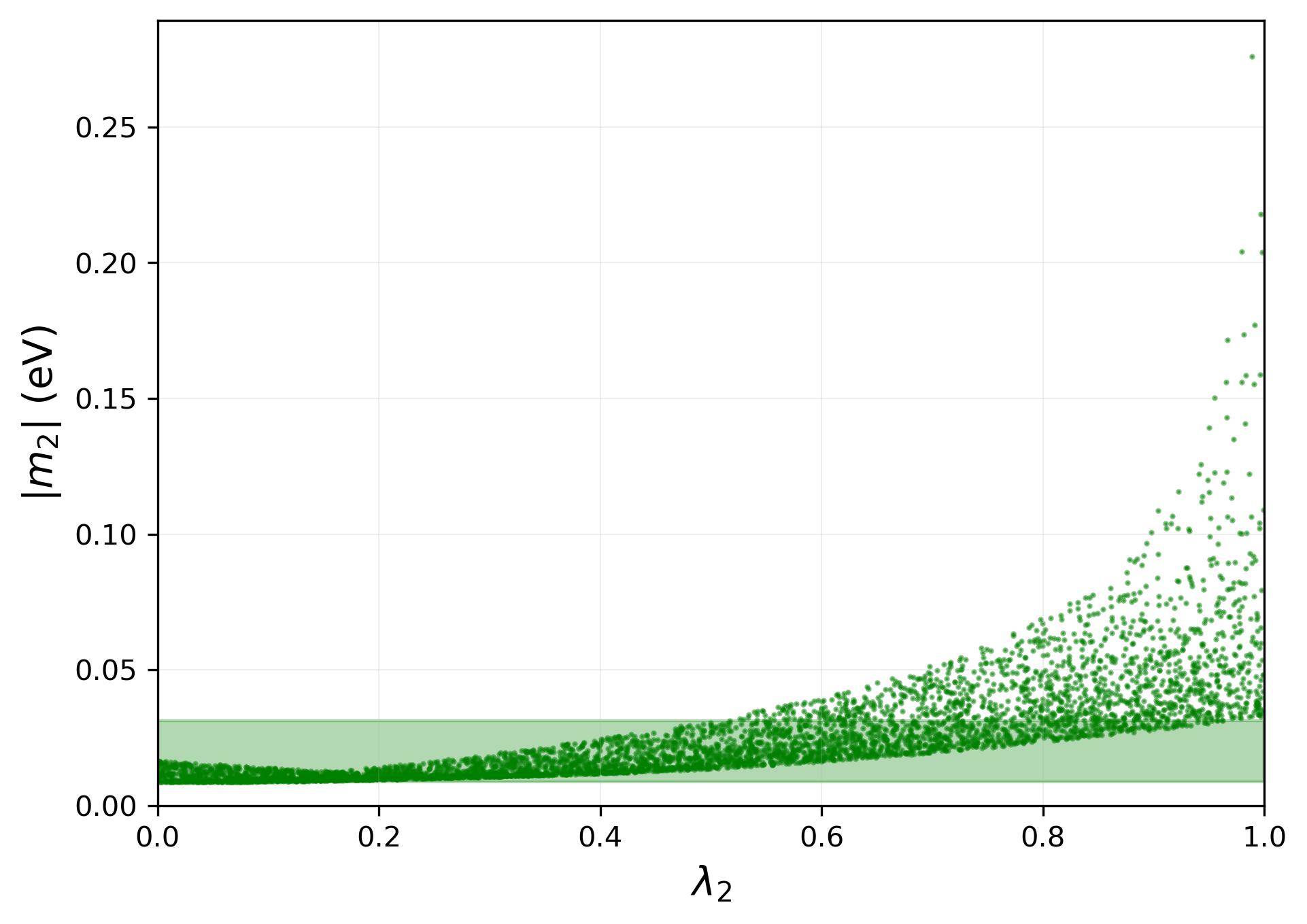}
    \caption{Correlation between the mass eigenvalue $|m_2|$ and $\lambda_2$ for the normal mass ordering (NO).}
    \label{fig:m2_vs_lambda2}
\end{figure}

\noindent
For all correlation plots in the normal ordering (NO) scenario, we adopt a common parameter setup in which $\theta = 145.3^\circ$ and $\psi = 181^\circ$ are fixed, while $\lambda_1$ and $\lambda_2$ are varied within the interval $[0,1]$.

Figure~\ref{fig:AC_correlation} shows the correlation between the mass eigenvalue $|m_2|$ and the neutrino mass matrix element $|(M_{\nu})_{11}|$ for NO scenario.
The light green horizontal band represents the allowed region for $|m_2|$ obtained from Eq.~(54).
For each randomly generated parameter point, the value of $|m_2|$ is computed using the model relation involving the solar mass-squared difference $\Delta m{21}^{2}$.



The correlations between $|m_2|$ and the parameters $\lambda_1$ and $\lambda_2$ for the NO scenario are shown in Figures~\ref{fig:m2_vs_lambda1} and~\ref{fig:m2_vs_lambda2}.
The mass--squared differences are taken at their best--fit values for normal ordering, $\Delta m_{21}^2 = 7.49 \times 10^{-5}~\text{eV}^2$ and $\Delta m_{32}^2 = 2.513 \times 10^{-3}~\text{eV}^2$.
The resulting correlations $|m_2|$ vs.\ $\lambda_1$ and $|m_2|$ vs.\ $\lambda_2$ are shown in the figures.
The light green horizontal band denotes the experimentally allowed range of $|m_2|$, as given in Eq.~(54).

\subsection{Inverted Order (IO) scenario}

We place special emphasis on the prediction of the Dirac CP phase 
$\delta$ near its maximal value of $270^\circ$, as indicated by the 
results of the T2K and NO$\nu$A experiments \cite{Abe2023T2K1, Abe2023T2K2, Himmel2020NOvA, Wolcott2024NOvA}. From the global analysis 
data presented in Table~I, we observe that this near-maximal value 
is particularly favored in the IO scenario. 
Furthermore, for such values of $\delta$ in the IO, 
the best-fit value of $\theta_{23}$ lies in the second octant. 
Motivated by this experimental indication, we therefore search 
for suitable ranges of the model parameters $\theta$ and $\psi$ 
that can simultaneously reproduce $\delta$ close to $270^\circ$ 
and $\theta_{23}$ in the second octant. From Fig.~\ref{fig:s23_theta_psi_0_21}, we observe that $\sin^2\theta_{23}$ lies in the second octant only for the $\theta$ range $124.2^\circ \leq \theta \leq 145.9^\circ$. Moreover, the minimal deviation of $\sin^2\theta_{23}$ from its maximal value is obtained for approximately 
$0^\circ \lesssim \psi \lesssim 5^\circ$. From Fig.~\ref{fig:s23_theta_psi_158_201}, we find that both allowed 
$\theta$ intervals, $34.1^\circ \leq \theta \leq 55.9^\circ$ and 
$124.2^\circ \leq \theta \leq 145.9^\circ$, yield $\sin^2\theta_{23}$ in the second octant, but for different approximate ranges of $\psi$. Specifically, the second-octant solution 
is realized for $180^\circ \leq \psi \leq 185^\circ$ in the first 
$\theta$ interval, and $175^\circ \leq \psi \leq 180^\circ$ in the 
second $\theta$ interval. Similarly, from Fig.~\ref{fig:s23_theta_psi_338_359}, we observe that 
the second octant is obtained for $34.1^\circ \leq \theta \leq 55.9^\circ$ when $355^\circ \leq \psi \leq 360^\circ$, as inferred from the color-gradient distribution in the plot.

\begin{figure}[t]
    \centering
    \includegraphics[width=0.60\textwidth]{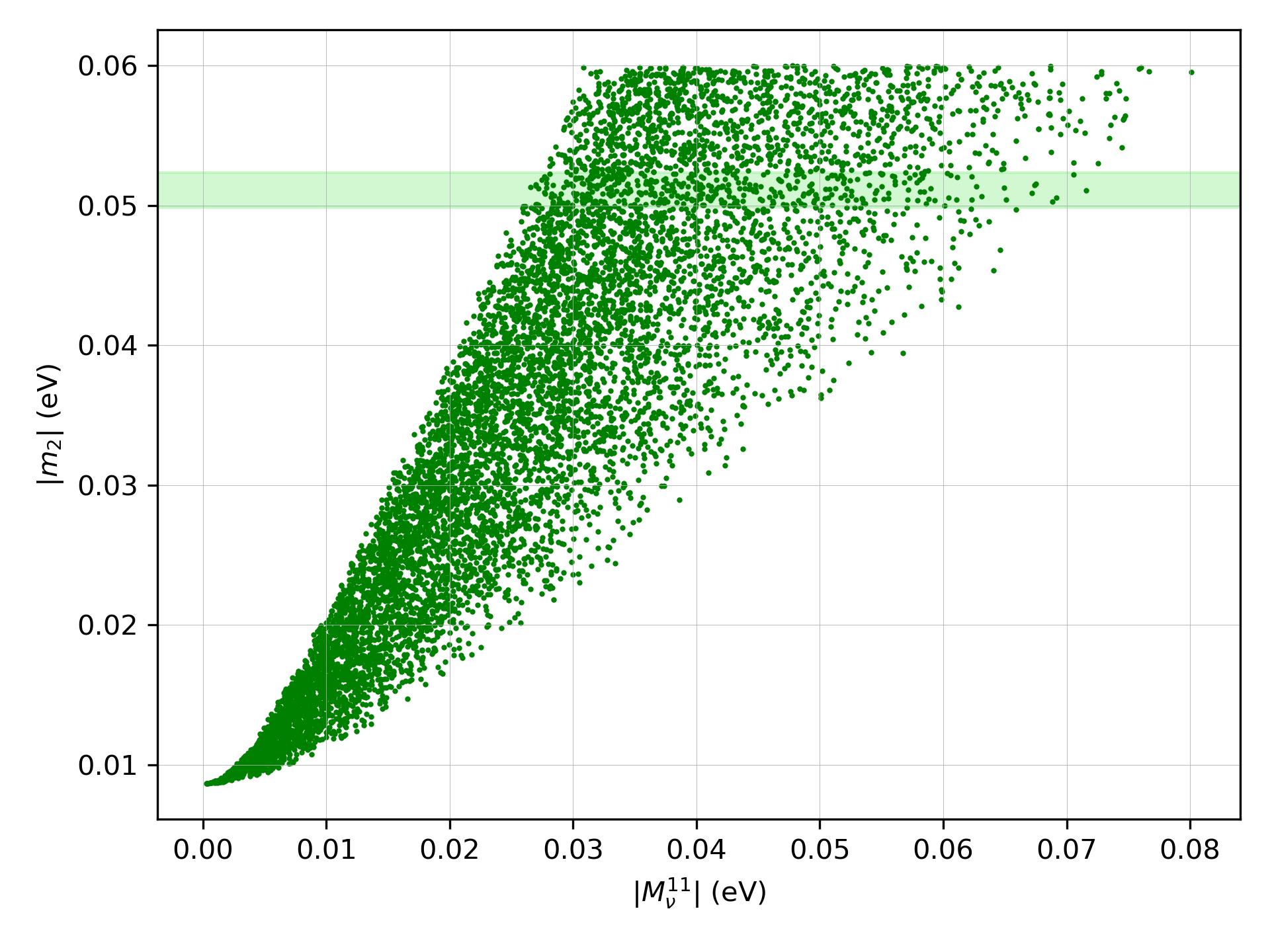}
    \caption{Correlation of the mass eigenvalue $|m_2|$ with the neutrino 
mass matrix element $|(M_{\nu})_{11}|$ in the inverted ordering (IO) case.}
    \label{fig:AC_correlationIO}
\end{figure}


In a similar manner, we now analyze the $\delta$--$\theta$ correlation plots to further constrain the allowed regions of the model parameters $\theta$ and $\psi$ for which $\delta$ deviates above $270^\circ$. From Fig.~\ref{fig:delta_theta_psi_0_21}, we observe that 
$\delta$ exceeds $270^\circ$ for the $\theta$ range 
$34.1^\circ \leq \theta \leq 45^\circ$ and for the approximate 
$\psi$ interval $0^\circ \leq \psi \leq 5^\circ$, 
as inferred from the color-gradient distribution in the plot.
From Fig.~\ref{fig:delta_theta_psi_158_201}, it is seen that 
$\delta > 270^\circ$ for the same $\theta$ range 
$34.1^\circ \leq \theta \leq 45^\circ$, but now for the approximate 
$\psi$ interval $180^\circ \leq \psi \leq 185^\circ$.
Similarly, from Fig.~\ref{fig:delta_theta_psi_338_359}, we find that 
$\delta$ exceeds $270^\circ$ for 
$34.1^\circ \leq \theta \leq 45^\circ$ when 
$355^\circ \leq \psi \leq 360^\circ$, 
again estimated from the corresponding color-gradient profile.

It is thus clear that the simultaneous realization of a second-octant solution of $\theta_{23}$ 
and $\delta > 270^\circ$ is achieved only within restricted regions 
of the parameter space. In particular, the allowed ranges are:
\begin{enumerate}
\renewcommand{\labelenumi}{(\roman{enumi})}
\item $34.1^\circ \leq \theta \leq 45^\circ$ and 
      $180^\circ \leq \psi \leq 185^\circ$,
\item $34.1^\circ \leq \theta \leq 45^\circ$ and 
      $355^\circ \leq \psi \leq 360^\circ$.
\end{enumerate}

We now determine the best-fit values of $\theta$ and $\psi$ within the above parameter ranges by using the reactor mixing angle, $\sin^{2}\theta_{13} = 0.02231$, obtained from the global analysis presented in Table~I for the IO case including SK data. First, we consider the allowed ranges $34.1^\circ \leq \theta \leq 45^\circ$ and $180^\circ \leq \psi \leq 185^\circ$. Within this region, the minimum value of $\sin^{2}\theta_{13}$ is approximately $0.6667$, which is much larger than the experimental best-fit value. This range is thus incompatible with the experimental data.
 Next, we consider the ranges $34.1^\circ \leq \theta \leq 45^\circ$ and $355^\circ \leq \psi \leq 360^\circ$. Within this region, $\sin^{2}\theta_{13}$ can attain values close to the experimental best-fit. A numerical evaluation yields $\theta \approx 34.8^\circ$ and $\psi \approx 356^\circ$, for which $\sin^{2}\theta_{13} \approx 0.02203$, in good agreement with the best-fit value $0.02231$. For this choice, we obtain $\sin^{2}\theta_{23} \approx 0.520$ and $\delta \approx 280.6^\circ$, which are reasonably close to their respective global best-fit values.
  Accordingly, we approximately fix the representative best-fit values of the model parameters as $\theta= 34.8^\circ$ and $\psi= 356^\circ$. These values are suitable for the IO scenario and lead to $\delta$ close $270^\circ$. They will be used in the subsequent numerical analysis.

\FloatBarrier
\begin{figure}[t]
    \centering
    \includegraphics[width=0.60\textwidth]{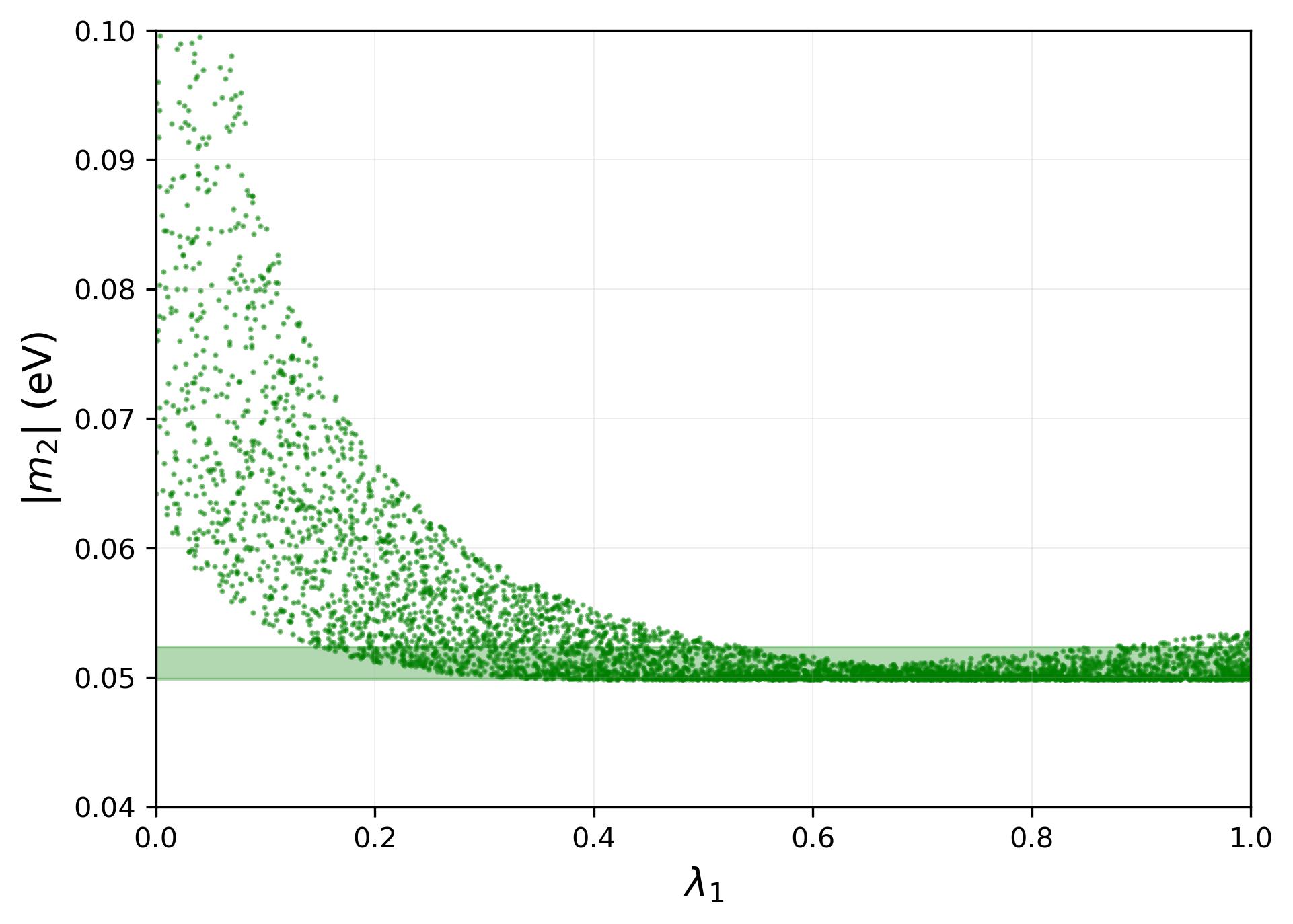}
    \caption{Correlation between the mass eigenvalue $|m_2|$ and $\lambda_1$ for the inverted mass ordering (IO).}
    \label{fig:m2_vs_lambda1IO}
\end{figure}



\begin{figure}[t]
    \centering
    \includegraphics[width=0.60\textwidth]{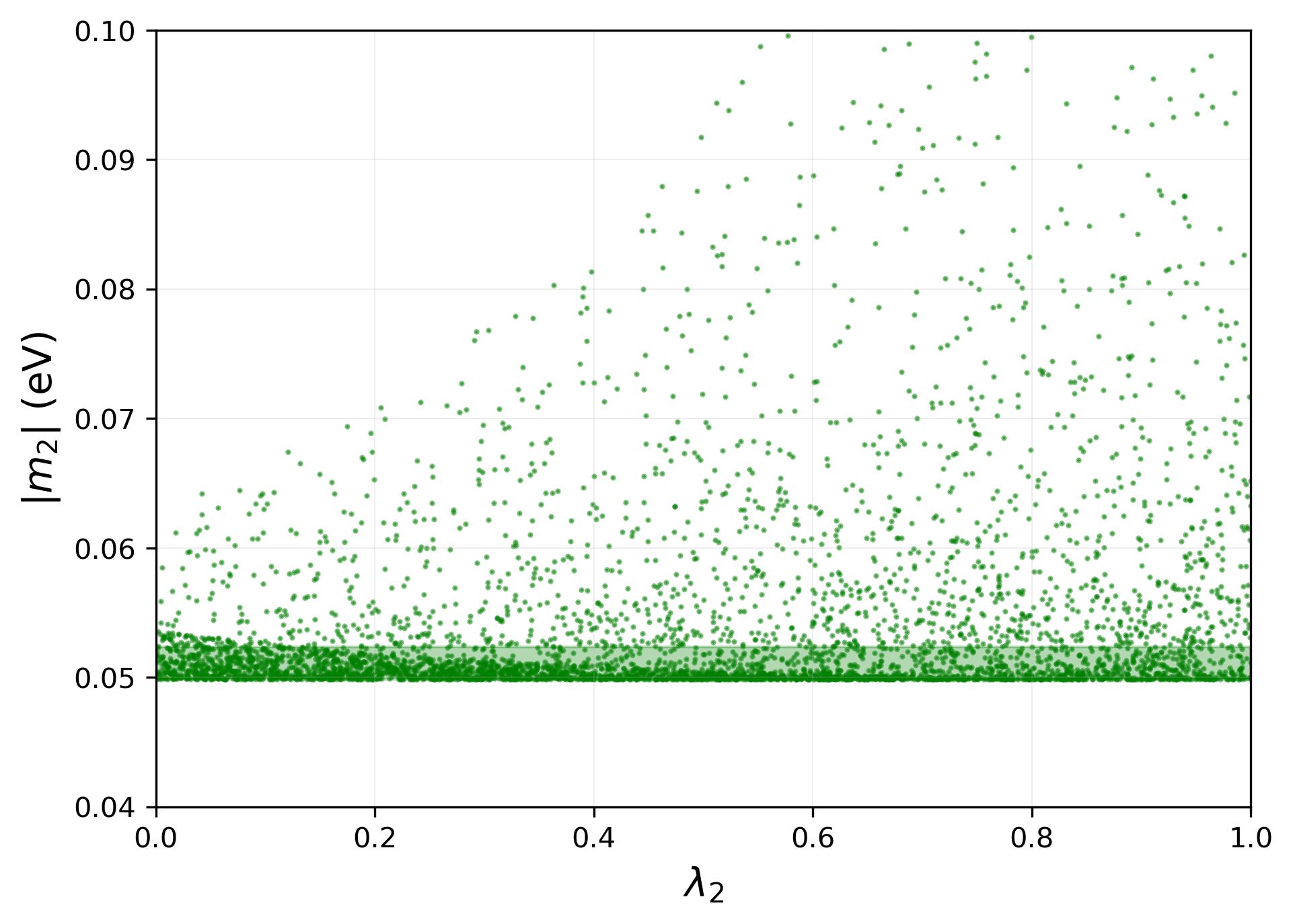}
    \caption{Correlation between the mass eigenvalue $|m_2|$ and $\lambda_2$ for the inverted mass ordering (IO).}
    \label{fig:m2_vs_lambda2IO}
\end{figure}

\FloatBarrier

Now we perform a correlation analysis among the mass matrix elements $|{M_\nu}^{11}|$, $|{M_\nu}^{22}|$, and the ratios of the mass matrix elements parameterized by $\lambda_1$ and $\lambda_2$, as we have done for the NO case. In a similar manner to the NO case, the allowed range of $|m_2|$ for the inverted ordering (IO) scenario, including the Super-Kamiokande (SK) data, is obtained as
\begin{equation}
|m_2|^{\min} \approx 0.049839~\text{eV},
\qquad
|m_2|^{\max} \approx 0.052350~\text{eV}.
\end{equation}

Fig.~\ref{fig:AC_correlationIO} illustrates how the neutrino mass 
matrix element $|(M_{\nu})_{11}|$ shows correlation with the mass eigenvalue 
$|m_2|$ in the inverted ordering (IO) scheme. 
The light green band marks the permitted range of $|m_2|$ obtained 
from Eq.~(65). In this analysis, $\lambda_{1}$ and $\lambda_{2}$ are 
scanned freely over the interval $[0,1]$. The mixing parameters are 
kept fixed at $\theta = 34.8^\circ$ and $\psi = 356^\circ$, values 
previously determined from experimental bounds on the mixing angles 
and the Dirac CP phase for the IO case. For every sampled point in 
the parameter space, $|m_2|$ is evaluated using the model relation 
that incorporates the solar mass-squared difference 
$\Delta m_{21}^{2}$.

For the inverted ordering case, we illustrate the dependence of $|m_2|$ on the parameters $\lambda_1$ and $\lambda_2$ in Figs.~\ref{fig:m2_vs_lambda1IO} and~\ref{fig:m2_vs_lambda2IO}. The parameters $\lambda_1$, $\lambda_2$, $\theta$, and $\psi$ are taken to be the same as those used in the above correlation analysis shown in Fig.~\ref{fig:AC_correlationIO}. The mass--squared differences are adopted at their best--fit values corresponding to the inverted ordering configuration, namely $\Delta m_{21}^2 = 7.49 \times 10^{-5},\text{eV}^2$ and $\Delta m_{32}^2 = -2.484 \times 10^{-3},\text{eV}^2$. The shaded horizontal strip (light green) indicates the phenomenologically allowed interval of $|m_2|$ given in Eq.~(55). The region where the scanned points intersect this band identifies the viable domain of $(\lambda_1,\lambda_2)$ compatible with current neutrino oscillation constraints.

\section{Summary and discussion}

The mixing pattern described by lepton mixing matrix exhibits a non-zero reactor angle, a well-measured solar angle, and an atmospheric mixing angle close to maximal. Current results from T2K and NO$\nu$A experiments also indicate a preference for the Dirac CP phase to lie near $3\pi/2$. These features suggest the presence of an underlying symmetry, among which $\mu$--$\tau$ reflection symmetry  emerges as a particularly appealing framework. Motivated by these observations, we construct a model based on the symmetry 
$SU(2)_L \times U(1)_Y \times A_4 \times Z_2 \times Z_4$ that yields a light
neutrino mass matrix realizing the $\mu$--$\tau$ reflection--symmetric texture. The model extends the scalar sector by introducing multiple
Higgs doublets and flavon fields, along with three right-handed neutrinos,
such that the light neutrino masses are generated via the Type-I seesaw mechanism. To obtain the $\mu$--$\tau$ reflection--symmetric mass matrix texture, we impose a generalized CP symmetry in the Yukawa Lagrangian. This renders all the Yukawa couplings and vevs real. In the symmetry limit, the lepton mixing matrix depends on a single parameter $\theta$. While $\theta_{23}$ and $\delta$ take maximal values, the remaining mixing angles, $\theta_{12}$ and $\theta_{13}$, are determined by $\theta$. 
To accommodate non-maximal values of $\theta_{23}$ and $\delta$, the neutrino mass matrix elements are allowed to be complex by relaxing the constraint of generalized CP symmetry imposed in the Yukawa Lagrangian. An additional parameter $\psi$ enters through the diagonalization of the light neutrino mass matrix. Thus, two parameters, $\theta$ and $\psi$, control the deviation from exact $\mu$--$\tau$ reflection symmetry and determine all the mixing angles and the Dirac CP phase.
A detailed numerical analysis has been performed by scanning the model parameters $\theta$ and $\psi$. Using the experimental constraints on $\theta_{12}$ and $\theta_{13}$, we identify the allowed regions of the parameter space. The results show that the allowed ranges are quite restricted, with $\theta$ confined to two distinct intervals, $34.1^\circ \leq \theta \leq 55.9^\circ$ and $124.2^\circ \leq \theta \leq 145.9^\circ$, while $\psi$ lies within three narrow intervals: $0.0^\circ \leq \psi \leq 21.8^\circ$, $158.2^\circ \leq \psi \leq 201.7^\circ$, and $338.3^\circ \leq \psi \leq 360^\circ$.

In the NO scenario, the parameter region that simultaneously yields $\sin^{2}\theta_{23}$ in the second octant and $\delta < 270^\circ$ is given by
$124.2^\circ \le \theta \le 145.9^\circ$ and $158.2^\circ \le \psi \le 180^\circ$. Within this allowed region, we identify representative values of the model parameters that provide good agreement with global-fit data. For instance, choosing $\theta = 145.2^\circ$ and $\psi = 179^\circ$, we obtain $\sin^{2}\theta_{13} \approx 0.0212$, which is consistent with its best-fit value $0.02195$. The corresponding prediction for the atmospheric mixing angle is $\sin^{2}\theta_{23} \approx 0.505$, lying in the second octant and reasonably close to the best-fit value $0.561$. Furthermore, the Dirac CP phase is predicted to be $\delta \approx 170^\circ$, in good agreement with the best-fit value $\delta \simeq 177^\circ$ from global analysis without SK data.
To satisfy the global results with the included SK data, the allowed parameter space is given by
$135^\circ \leq \theta \leq 145.9^\circ$ and $180^\circ \lesssim \psi \lesssim 200^\circ$,
which yield the first-octant solution for $\theta_{23}$ and a Dirac CP phase close to $\delta \simeq 212^\circ$.
Within this region, we choose $\theta \approx 145.3^\circ$ and $\psi \approx 181^\circ$ as representative values such that the model predicts $\sin^{2}\theta_{13} \approx 0.0214$, in good agreement with the best-fit value $\sin^{2}\theta_{13} = 0.02215$ for the NO scenario including SK data. The corresponding atmospheric mixing angle is $\sin^{2}\theta_{23} \approx 0.494$, which lies in the first octant and is reasonably close to its best-fit value $0.470$. Furthermore, the Dirac CP phase is predicted to be $\delta \approx 226^\circ$, which is fairly close to the global best-fit value $\delta \simeq 212^\circ$. In the IO scenario, compatibility with $\sin^{2}\theta_{23}$ in the second octant and $\delta$ close to $270^\circ$ is achieved for the parameter ranges
$34.1^\circ \leq \theta \leq 45^\circ$ and $355^\circ \leq \psi \leq 360^\circ$. A numerical analysis yields representative values $\theta \approx 34.8^\circ$ and $\psi \approx 356^\circ$, for which $\sin^{2}\theta_{13} \approx 0.02203$, in good agreement with the best-fit value $0.02231$. The corresponding prediction for the atmospheric mixing angle is $\sin^{2}\theta_{23} \approx 0.520$, lying in the second octant, while the Dirac CP phase is $\delta \approx 280.6^\circ$. Both predictions are reasonably close to their respective global best-fit values.

In conclusion, the $A_4$ model presented in this work can accommodate maximal $\theta_{23}$ and maximal $\delta$ in a generalized CP symmetry limit. In the general situation (without invoking the generalized CP symmetry), it can predict deviations from maximality consistent with global analysis data.
The constraints on the allowed regions of the model parameters will depend on precise measurements of the lepton mixing parameters.

\appendix

\section*{Appendix A}
\subsection*{Basic $A_{4}$ properties}

The alternating group of degree four, denoted $A_{4}$, consists of the 
twelve even permutations of four objects. It is isomorphic to the rotational 
symmetry group of a regular tetrahedron. The irreducible representations are 
one triplet $\mathbf{3}$ and three singlets $\mathbf{1}$, $\mathbf{1}'$, 
$\mathbf{1}''$.

The tensor product of two triplets is given by
\begin{equation}
\mathbf{3}\otimes \mathbf{3} = 
\mathbf{3}_{s}\oplus \mathbf{3}_{a}\oplus 
\mathbf{1}\oplus \mathbf{1}'\oplus \mathbf{1}''.
\end{equation}

The singlet multiplication rules are
\begin{equation}
\mathbf{1}'\otimes \mathbf{1}' = \mathbf{1}'', \quad 
\mathbf{1}'\otimes \mathbf{1}'' = \mathbf{1}, \quad 
\mathbf{1}''\otimes \mathbf{1}'' = \mathbf{1}'.
\end{equation}

For two triplets $(a_{1},a_{2},a_{3})$ and $(b_{1},b_{2},b_{3})$, the decomposition is
\begin{align}
(\mathbf{3}\otimes \mathbf{3})_{\mathbf{3}_{s}} 
&= (a_{2}b_{3}+a_{3}b_{2},\; 
    a_{3}b_{1}+a_{1}b_{3},\; 
    a_{1}b_{2}+a_{2}b_{1}), \\
(\mathbf{3}\otimes \mathbf{3})_{\mathbf{3}_{a}} 
&= (a_{2}b_{3}-a_{3}b_{2},\; 
    a_{3}b_{1}-a_{1}b_{3},\; 
    a_{1}b_{2}-a_{2}b_{1}), \\
(\mathbf{3}\otimes \mathbf{3})_{\mathbf{1}} 
&= a_{1}b_{1}+a_{2}b_{2}+a_{3}b_{3}, \\
(\mathbf{3}\otimes \mathbf{3})_{\mathbf{1}'} 
&= a_{1}b_{1}+\omega a_{2}b_{2}+\omega^{2}a_{3}b_{3}, \\
(\mathbf{3}\otimes \mathbf{3})_{\mathbf{1}''} 
&= a_{1}b_{1}+\omega^{2}a_{2}b_{2}+\omega a_{3}b_{3},
\end{align}
where $\omega = e^{2\pi i/3}$.

\end{document}